\documentstyle[psfig]{mn2e}

\newif\ifAMStwofonts

\def\simlt{\lower.5ex\hbox{$\; \buildrel < \over \sim \;$}}
\def\simgt{\lower.5ex\hbox{$\; \buildrel > \over \sim \;$}}

\def\mnras{MNRAS}
\def\apj{ApJ}
\def\apjl{ApJL}
\def\aap{AAp}
\def\pasj{PASJ}

\title[The Evolution of Substructure III]
    {The Evolution of Substructure in Galaxy, Group \\ and Cluster Haloes III:
Comparison with Simulations}
\author[Taylor \& Babul]
{James E. Taylor$^{1}$\thanks{PPARC Fellow}\thanks{email: {\tt jet@astro.ox.ac.uk}} and Arif Babul$^{2}$ \\
$^{1}$Denys Wilkinson Building, 1 Keble Road, Oxford OX1 3RH, United Kingdom \\
$^{2}$Elliott Building, 3800 Finnerty Road, Victoria, BC, V8P 1A1, Canada \\}
\date{\today}
\pubyear{2004}

\begin{document}

\maketitle

\begin{abstract}
In a previous paper, we described a new method for including detailed
information about substructure in semi-analytic models of halo
formation based on merger trees. In this paper, we compare the
predictions of our model with results from self-consistent numerical
simulations. We find that in general the two methods agree extremely
well, particularly once numerical effects and selection effects in
the choice of haloes are taken into account. As expected from the
original analyses of the simulations, we see some evidence for
artificial overmerging in the innermost regions of the simulated
haloes, either because substructure is being disrupted artificially
or because the group-finding algorithms used to identify substructure
are not detecting all the bound clumps in the highest-density
regions. Our analytic results suggest that greater mass and force
resolution may be required before numerical overmerging becomes
negligible in all current applications. We discuss the implications
of this result for observational and experimental tests of halo
substructure, such as the analysis of discrepant magnification ratios
in strongly lensed systems, terrestrial experiments to detect dark
matter particles directly, or indirect detection experiments searching
for positrons, gamma-rays, neutrinos or other dark matter decay
products.
\end{abstract}



\section{Introduction}\label{sec:1}

There is now very strong evidence from observations of the microwave 
background (Spergel et al.\ 2003), galaxy redshift surveys (e.g.\ 
Tegmark et al.\ 2004), weak lensing measurements (e.g.\ Rhodes et al.\ 
2004), and modelling of the Lyman-$\alpha$ forest (e.g.\ Kim et al.\ 2004),
that most of the matter in the universe is non-baryonic dark matter,
and that the power spectrum of density fluctuations in this component 
extends to subgalactic scales, as expected in `cold' dark matter (CDM) 
models. The implications of the CDM power spectrum for structure formation 
are well established. Dark matter haloes, the dense regions that surround
galaxies, groups and clusters, form from the bottom up, through the
merging of progressively larger structures. This process of hierarchical 
merging has been studied extensively, and the overall properties of galaxy 
or cluster haloes formed in this way are now fairly well determined.

To learn more about dark matter, and to search for features in the
power spectrum that could reveal new phases in the evolution of the
very early universe, we must push the theory of structure formation
to smaller scales. Most of our current understanding of the properties 
of dark matter on subgalactic scales comes from numerical simulations 
of structure formation. These simulations have been used to determine 
the evolution of large-scale structure and the formation of CDM haloes 
on scales ranging from the current horizon 
(Kauffmann et al.\ 1999) down to the local neighbourhood 
(e.g\ Mathis et al.\ 2002). Furthermore, by selectively re-simulating 
sections of a large volume at higher resolution, recent studies have been 
able to `zoom in' on single objects, resolving the substructure within 
individual haloes in exquisite detail (e.g.\ recent work by 
De Lucia et al.\ 2004; Gill, Knebe, \& Gibson 2004a; Gill et al.\ 2004b; 
Gao et al.\ 2004a, 2004b; Diemand et al.\ 2004c; 
Weller, Ostriker \& Bode 2004; Reed et al.\ 2004) 

There is a hard limit, however, to the dynamic range that can be
achieved using  this approach of selective re-simulation. Structure
formation mixes information  on many different scales as haloes
form. To model the formation of a dark matter halo accurately, one
needs to include the effects of very long-wavelength fluctuations as
well as the smaller fluctuations that produce substructure. The
minimum scale that can be included in any self-consistent  simulation
of the formation of a present-day halo is determined by the
requirement that the largest fluctuations in the volume studied still
be  in the linear regime at the present day, and by the finite
numerical  resolution available computationally. For the
highest-resolution simulations  that are currently feasible, this
leads to a minimum mass scale for resolved  substructure of around
10$^{-4}$--10$^{-5}$ of the mass of the main halo  considered. To
study halo substructure below this mass limit requires analytic  or
semi-analytic extensions to the numerical results. It is precisely
this sort  of small-scale information, however, that is required in
many current  applications including galaxy dynamics, strong lensing,
direct or indirect  dark matter detection, or tests of dark matter
physics in general.

In earlier work (Taylor \& Babul 2001, TB01 hereafter), we developed a
model for dynamical evolution of satellites orbiting in the potential
of larger system. This model includes simple treatments of dynamical
friction, tidal mass loss and tidal disruption.  It calculates
satellite evolution over a many short timesteps, rather like a
restricted $N$-body simulation, but uses only global properties of the
satellite to determine its evolution, thus reducing the computational
expense considerably.  More recently (Taylor \& Babul 2004a, paper I
hereafter), we have applied  this model of satellite evolution to the
merging subcomponents involved in  the hierarchical formation of
galaxy, group or cluster haloes, creating a  full semi-analytic model
of halo formation. In a second paper  (Taylor \& Babul 2004b, paper II
hereafter), we presented the basic predictions  of this model,
including distributions of subhalo mass, circular velocity,  location
and merger epoch, and the correlations between these properties.  We
found results similar to those of recent numerical studies, as well as
for a few systematic differences.

In this paper, we compare the predictions of the semi-analytic model
directly with the results of self-consistent numerical simulations of
halo formation. This comparison is particularly interesting, since the
only free parameters in the semi-analytic model were fixed in paper I,
either by matching restricted simulations of individual subhaloes (to
fix the parameters of the dynamical model), or by assuming
self-similarity in the merging process (to fix the one free parameter
in the pruning method). Thus we have no remaining parametric freedom
when comparing our results to self-consistent simulations, making the
comparison a meaningful one.  Overall, we will show that there is
reasonable agreement between the  semi-analytic  and numerical
results, particularly in regions where both are expected to be
accurate, but also that there are systematic differences between their
predictions. These could reflect inaccuracies in the semi-analytic
model,  but closer examination of the numerical results suggests that
at least part of the discrepancy is due to artificial numerical
effects in the simulations. The quantitative estimate of the magnitude
of these effects has interesting implications for the analysis of
several recent observational results.

The outline of this paper is as follows. In section \ref{sec:2}, we
summarise briefly the semi-analytic model developed in paper I. In
section \ref{sec:3}, we describe the six simulated haloes used in our
comparison, and analyse the general properties of their subhalo
populations. In section \ref{sec:4}, we compare the properties of
individual subhaloes, as well as the cumulative distributions of
subhalo mass or circular velocity, in semi-analytic model and in the
numerical simulations. In particular, we examine the evidence that
the central regions of the simulated haloes are subject to
artificial overmerging. In section \ref{sec:5}, we consider the
implications of overmerging in two particular areas, the modelling
of strongly-lensed systems, and the analysis of direct detection
experiments. We summarise our conclusions in section \ref{sec:6}. 
Finally, we note that as in papers I and II, in this paper we will 
generally consider results for the former `standard' CDM (SCDM) 
cosmology with $h = 0.5$ and $\sigma_8 = 0.7$, because the simulations 
we compare to assumed this cosmology. In general, our main results 
depend only weakly on cosmology, as discussed in paper II.

\section{Review of the semi-analytic model}\label{sec:2}

In paper I, we introduced a semi-analytic model for studying the
formation of dark matter haloes and the evolution of their
substructure. In this section we will review briefly the main features
of this model. The model is explained fully in TB01 and paper I, and
a more detailed summary is given in paper II.

The semi-analytic model consists of several components: a method for
generating merger trees, an algorithm for `pruning' these trees, to
determine how many distinct satellites merge into the main system
within the tree, an analytic model to describe the subsequent
evolution of these satellites, and a model for the concurrent
evolution of the main system. The halo merger histories are generated
using the merger-tree algorithm of Somerville and Kolatt
(1999). Higher order branchings in these trees are then pruned, using
the method described in paper I, to determine whether each branch
merging with the main trunk contributes a single subhalo or a group
of associated subhaloes to the main system. This produces a single
list of subhaloes merging with the main system at various
redshifts. Each subhalo from this final list is the placed on a random
orbit starting at the virial radius of the main system, and evolved
using the analytic model of satellite dynamics described in TB01,
experiencing orbital decay due to dynamical friction, and heating and
stripping due to tidal forces. Haloes which were associated with a
given parent before pruning fall in together with the parent on
similar orbits, as part of a kinematic group. 

The properties of the main system also change over time, its mass growing 
according to the merger tree and its concentration changing according to 
the relations in Eke, Navarro, \& Steinmetz (2001, ENS01 hereafter). 
Although no baryonic component is included in the 
models presented here, one can easily be added, given a prescription for gas
cooling and star formation. We assume, unless specified otherwise, that the 
main system has a Moore density profile and a concentration or scale
radius given by the relations in ENS01. Our fiducial system, a
$1.6\times10^{12}M_{\odot}$ halo at $z = 0$ in a SCDM cosmology, has a
concentration $c_{\rm M} = 10.3$, a scale radius $r_{\rm s,M} =
30.5$\,kpc, a virial radius $r_{\rm vir,m} = 314.1$\,kpc, and a
virial velocity (or circular velocity at the virial radius) 
$v_{\rm vir,m} = 148$\,km\,s$^{-1}$. We note that this concentration
is typical for a galaxy of this mass (ENS01); galaxy clusters would
be about half as concentrated, this difference should be kept in mind
when comparing our results with simulations of more massive systems. 
On the other hand, real galaxy haloes have large concentrations of
baryonic material at their centres, and through adiabatic contraction 
they may have become more concentrated than the systems considered here; 
this possible difference should be kept in mind when comparing with
observations. 

In all, the dynamical model has two main free parameters -- the Coulomb
logarithm $\Lambda_{\rm s}$ which modulates dynamical friction, and
the heating coefficient $\epsilon_{\rm h}$ which modulates mass
loss. (A third parameter discussed in TB01, the disk logarithm
$\Lambda_{\rm d}$, is not used here since we are considering
evolution in a single-component potential). The precise disruption
criterion (say the fraction of the binding radius used to define
$f_{\rm dis}$), the form chosen for the density profile of the
satellites and the profile of the main system, and various other model
choices will also affect some of our results, though not very
strongly. We discuss the model-dependence of our results in
paper II. Here we generally present results for the default
parameter values discussed in paper I, specifically 
$\Lambda_{\rm s} = 2.4$ (where the magnitude of dynamical friction 
scales as $\Lambda (M) = \Lambda_{\rm s} + \ln(M_{\rm h}/140\,M_{\rm s})$ 
if $m < M/140$, and $\Lambda (M) = \Lambda_{\rm s}$ for $m \ge M/140$), 
and $\epsilon_{\rm h} = 3.0$. The disruption criterion assumes either
$f_{\rm dis} = 0.5$ (model A) or $f_{\rm dis} = 0.1$ (model B). Given
these parameter choices, the pruning parameters are fixed iteratively
as discussed in paper I. 

\section{Numerical Predictions for Halo Substructure}\label{sec:3}

\subsection{Review of the simulations}\label{subsec:3.1}

To test the accuracy of our model and compare it with fully numerical
results, we will examine the properties of substructure in six
different haloes extracted from high-resolution simulations. The
basic properties of these haloes are listed in table 1, along with the
references
in which the original simulations are described. The subhalo lists
extracted from these simulations were supplied by their respective
authors; in some cases they differ slightly from the data sets used in
the references listed, as the simulations have been reanalysed
subsequently. We will start by examining these datasets in detail, to
quantify how much scatter is expected in subhalo properties from one
system to another. We note that a much larger sample of $\Lambda$CDM
haloes, simulated at comparable or higher resolution, has recently
become available (De Lucia et al.\ 2004; Desai et al.\ 2004; 
Gill et al.\ 2004a, 2004b; Diemand et al.\ 2004c; Gao et al.\ 2004a, 2004b; 
Weller et al.\ 2004; Reed et al.\ 2004). Wherever possible, we
will also consider this more recent work.

The objects named `Coma' and `Virgo I' are a massive and an
intermediate-mass cluster halo, respectively, extracted from the
simulations of Moore et al.\ 1998 (M98 hereafter). `Virgo IIa' and
`Virgo IIb' are actually two different outputs from the same
simulation of a Virgo-sized cluster, at redshifts 0 and 0.1
respectively. The cluster, described in Ghigna et al.\ 2000 (G00), is
a higher-resolution re-simulation of a system first discussed in
Ghigna et al.\ 1998 (G98). `Andromeda' and the `Milky Way' (the
`Local Group') are a close pair of galaxy-size haloes selected
because of their resemblance to the two main systems in the real Local
Group. They are described in Moore et al.\ 1999b (M99b) and their
substructure is analysed in Moore et al.\ 1999a (M99a).

These simulations were all performed in a `standard' CDM 
($\Omega = 1$, $h = 0.5$, $\sigma_8 = 0.7$) cosmology. For purposes of
comparison, we have generated our semi-analytic results assuming the
same cosmology. The simulations cover a wide range of mass, and also
a range in mass resolution and softening length, as indicated in
table 1.
They typically have several million particles within the virial
radius, and a softening length of less than 1 percent of the virial
radius. Although these simulations were performed several years ago,
this combination of mass and force resolution has only recently been
surpassed more than a factor of 1.5--2, and even then only in a very
few simulations (e.g.\ Diemand et al.\ 2004c; Gao et al.\ 2004b).
Virgo IIa and IIb have particularly high force resolution, as well as
their high mass resolution. Coma has comparable mass resolution but
more softening, while Virgo I, Andromeda and the Milky Way have lower
mass resolution, and are also more heavily softened.

The substructure in these simulations was identified using the group
finder SKID (Stadel 2001; available at http://
hpcc.astro.washington.edu/tools). SKID identifies groups by finding
local maxima in the density field, linking them together with a
friends-of-friends algorithm, and then removing unbound particles
iteratively. It produces estimates of the structural properties of
each bound group of particles, including its total mass, its outer
radius (the radius of the outermost bound particle), the radius at
which its rotation curve peaks, and the value of the peak circular
velocity. We have all of this information for the subhaloes in the
Virgo II and Local Group simulations, and more limited information
for the first two simulations. Of the various properties measured by
SKID, we will assume that the total mass $M_{\rm s}$ is slightly more
reliable than the outer radius, since the latter depends on the
position of the single outermost particle. We will also consider the
peak velocity $v_{\rm p,s}$ of each subhalo, as an indicator of its
density profile and concentration.

We note that the structural properties of individual subhaloes in
simulations are subject to important numerical effects. This has been
demonstrated by carrying out idealised simulations of satellites in
fixed potentials, at much higher resolution than is possible in
self-consistent simulations where haloes form naturally from
cosmological initial conditions (Hayashi et al.\ 2003, H03 hereafter; 
Kazantzidis et al.\ 2004). Even
in a static potential, determining rotation curves for subhaloes to
an accuracy of 10 percent after a few orbits requires resolving them
with more than $5\times 10^5$ particles initially (i.e. a few times
$10^4$ after mass loss -- Kazantzidis et al.\ 2004). Given the
steepness of the cumulative velocity function, a 10 percent error in
velocity can change the number of subhaloes at a given velocity by
30--40 percent, so even errors of this order should be taken into
account. Force softening also has a direct effect on small subhaloes,
placing an upper limit on their circular velocity when they are
sufficiently dense. Finally, the group-finding algorithms used to
identify substructure in self-consistent simulations often depend
explicitly on the local density of a subhalo's environment. Thus
subhalo properties should be treated with caution even in
high-resolution simulations. We will discuss these issues further in
sections \ref{subsec:4.1} and \ref{subsec:4.2} below.

Finally, we need to normalise the properties of each set of numerical
subhaloes, in order to compare them on the same footing. To do so, we
divide the mass of each subhalo by $M_{\rm vir,m}$, the mass of its
parent halo within its virial radius, and divide the peak velocity of
the subhalo by $v_{\rm vir,m}$, the circular velocity of its parent
halo at the virial radius\footnote {In the case of the `Local Group'
haloes, the mass of the main halo was measured at $z = 0$, whereas
our outputs are for $z = 0.2$. We have assumed that the halo masses
were 0.885 of their final value at this redshift, based on the
average accretion rate measured in our merger trees. The virial
radius for an object of a given mass is also smaller at $z > 0$,
since it is defined in terms of a fixed overdensity relative to the
background.}. Where necessary we can then scale these relative values
to our semi-analytic model values, multiplying them by $M =
1.6\times\,10^{12}\,M_{\odot}$ and 148\,km\,s$^{-1}$, respectively.
When counting the number of subhaloes over some mass or velocity
threshold, we generally limit ourselves to the region within the
virial radius of the main halo, since the semi-analytic results are
incomplete beyond the virial radius, as they do not include subhaloes
that have not yet fallen in past this point. This procedure produces
relative distributions or scaled distributions that can easily be
compared with one-another and with the semi-analytic
results. Furthermore, we expect the properties of each system to be
similar when scaled in this way, since structure formation should be
fairly close to scale-invariant over the range of halo masses
considered here.

\subsection{Scatter in the numerical distributions}\label{subsec:3.2}

\begin{figure}
 \centerline{\psfig{figure=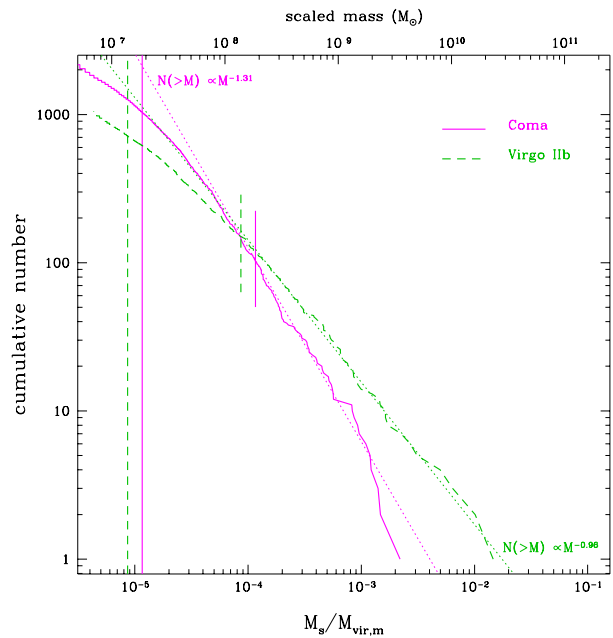,width=1.0\linewidth,clip=,angle=0}}
 \caption[]{The cumulative relative mass functions for the two
highest-resolution simulations, Coma (solid line) and Virgo II (dashed
line; shown at z = 0.1). The dotted lines are power-laws with slopes
-1.31 and -0.96. The top axis shows the equivalent subhalo mass in a
system with the fiducial mass $1.6\times 10^{12} M_{\odot}$. The
vertical lines indicate the 32-particle and 320-particle
mass-resolution limits for each simulation. 
}\label{fig:0}
\end{figure}

\subsubsection{The shape of the mass function}

Fig.\ \ref{fig:0} shows the cumulative relative mass functions for
all subhaloes within the virial radius of two systems, Coma (solid
line) and Virgo IIb (dashed line). The full vertical lines indicate
the mass for each simulation corresponding to 32 particles. In the
original analysis of the simulations, this was generally chosen as the
limit below which the results from the group finder became
significantly incomplete, and the structural parameters of subhaloes
unreliable. In fact we expect resolution effects to remain important
at much larger masses. As discussed in Diemand et al.\ (2004a), the
mean relaxation time for cuspy systems with density profiles similar
to those of subhaloes is less than a Hubble time when they are
resolved with a few hundred or even a thousand
particles. Furthermore, this calculation assumes present-day
densities (e.g. a half-mass radius of 24\,kpc for a system of mass
$3.5\times 10^9 M_{\odot}$, versus $\sim$20\,kpc for an isolated halo
of the same mass at $z = 0$ in our model). For systems which formed
at redshift $z$ the relaxation time should be shorter by a factor 
$(1 + z)^{3/2}$. Thus we also include shorter lines showing a 320-particle
mass, below which most systems should be artificially relaxed.

\begin{figure}
 \centerline{\psfig{figure=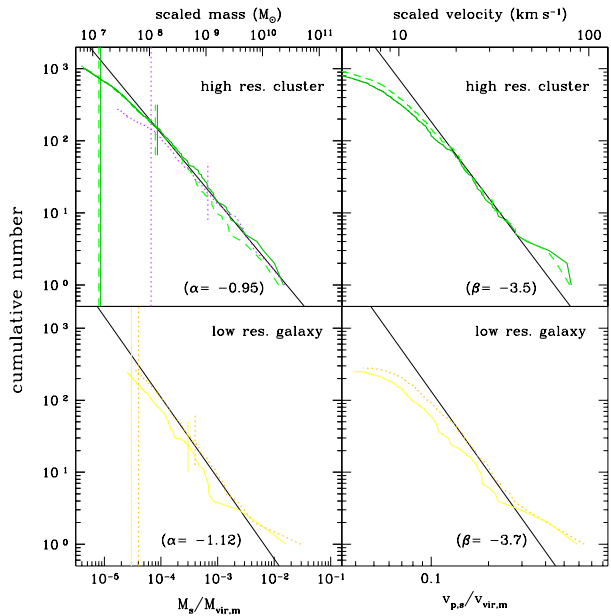,width=1.0\linewidth,clip=,angle=0}}
 \caption[]{Cumulative relative mass and velocity distributions from
various simulations of haloes on cluster (top panels) and galaxy
(bottom panels) scales. The top left-hand panels show the cumulative
mass functions for Virgo I (dotted line) and Virgo IIa \& IIb (solid
and dashed lines), while the bottom left-hand panel shows the relative
mass function for Andromeda (solid line) and the Milky Way (dotted
line). The bottom axis shows mass relative to the virial mass, while
the top axis shows the mass scaled to our fiducial halo. A line of
slope $\alpha$ is also shown on each plot. The vertical lines indicate
the 32-particle and 320-particle mass-resolution limits for each
simulation, and the right-hand panels show cumulative distributions
of peak circular velocity for Virgo IIa \& IIb (top right) and for
the Local Group (bottom right), as well as lines of slope $\beta$.
The bottom axis gives the value relative to the virial velocity of the
main halo, while the top axis gives the velocity scaled to our
fiducial system. }\label{fig:1}
\end{figure}

In each case, the cumulative mass function is roughly a power-law at
intermediate masses. The slope of the mass function differs
substantially between the two simulations, however -- it is about
$-0.96$ for Virgo IIb and $-1.31$ for Coma, as indicated by the dotted
lines. A priori, it is not clear whether this difference is due to
intrinsic, halo-to-halo variation in the mass function, the different
masses of the two systems, their redshifts or their different
internal dynamical states, or whether it is the result of different
softening and mass resolution. The latter seems unlikely given the
large difference even for well-resolved ($10^3$--$10^4$)
subhaloes. From the discussion in paper II, it seems
likely that dynamical age is an important factor. The progenitor of
the `Coma' halo formed in isolation and was fully relaxed at $z = 0$
(M98), while Virgo IIb, at a redshift of 0.1, contains massive
subsystems that have not yet been stripped or disrupted to the same
degree.

We also see in this figure that at low masses, the cumulative mass
function deviates from a power-law well before the 32-particle limit
of the group-finder is reached, but somewhere in the 100--300
particle range below which relaxation may be important. Here again,
though, it is not clear how much of the curvature of the mass
function is real and how much is numerical. We will discuss this
further when we compare these results to the semi-analytic predictions.

\subsubsection{Dependence on halo mass}

We can test whether the mass function depends on halo mass in a simple
way by comparing results for galaxy and cluster haloes. Fig.\
\ref{fig:1} shows the relative mass functions for all the
simulations of Virgo-sized haloes (top left panel), as well as the
mass functions for the two galaxy-sized haloes (bottom left panel).
The top axis indicates the equivalent subhalo mass and circular
velocity in our fiducial system (i.e. for 
$M_{\rm vir,m} = 1.6\times 10^{12} M_{\odot}$). The vertical lines 
indicate a 32-particle and a
320-particle lower mass limit in each simulation.

All five scaled mass functions are similar, although both the
normalisation and the slope vary by $\sim$ 20 percent. The variation
in normalisation depends on our convention for rescaling the
distributions; if we were to count all the haloes within 
$1.5\,r_{\rm vir,m}$, for instance, then Andromeda would have more 
satellites than the Milky Way. We will only count subhaloes within 
$1.0\,r_{\rm vir,m}$, however, as the semi-analytic results are 
incomplete beyond this point, as explained previously.

The variation in slope is also hard to define precisely, since the
mass functions deviate from a power-law at both large and small
masses, either for physical reasons or for numerical ones. Still,
there is a significant difference between the two sets of mass
functions. The thin solid lines show a rough fit to the slope of the
mass function at intermediate mass, with the logarithmic slope
$\alpha$ indicated on the plot. The trend in the slope going from
galaxies to clusters is the opposite of the one in Fig.\
\ref{fig:0}, in the sense that the less massive systems have steeper
mass functions, so it cannot be explained simply in terms of halo
mass. Instead it may reflect the dynamical ages of the different
systems, as discussed above and in paper II. In this
case, the Local Group haloes would be systematically older than Virgo,
just as Coma is.

Similar results have been reported recently for $\Lambda$-CDM
simulations. De Lucia et al.\ (2004), for instance, find logarithmic
slopes of $-0.97$ to $-0.98$ for the mass function on cluster scales,
and $-1.11$ to $-1.13$ on galaxy scales (although the quantity they
fit is ${\rm d}n(M)/{\rm d}(\log(M))$ versus $\log(M)$, for power law
distributions the slope of this quantity has the same numerical value
as the logarithmic slope of the cumulative distribution
$\alpha$). The trend to steeper slopes for smaller haloes is as in
Fig.\ \ref{fig:1}.

\subsubsection{Shape of the cumulative velocity distribution}

The right-hand panels of Fig.\ \ref{fig:1} show the cumulative
distributions of peak velocity, either relative to the virial velocity
of the main system (bottom axis), or scaled by our fiducial value of
148\,km\,s$^{-1}$ (top axis). These are also well described by
power-laws at intermediate mass, as indicated by the thin solid
lines. The logarithmic slope $\beta$ is indicated on the plot. For
self-similar haloes we expect $v_{\rm p} \propto M^{1/3}$ and
therefore $\beta = 3\alpha$; in practice the slope seems slightly
steeper than this, perhaps indicating that the small subhaloes are
more concentrated than the large ones. The velocity distributions
show stronger deviations from a power-law at small velocities than 
the mass functions do at low masses; we will discuss a possible
explanation for this in section \ref{subsec:4.2}.

Finally, we note that these mass and velocity distributions are
similar to, and consistent with, others that have appeared in the
literature (e.g.\ Klypin et al.\ 1999b; Okamoto \& Habe 1999; Springel
et al.\ 2001; Governato et al.\ 2001; Stoehr et al.\ 2002; De Lucia et
al.\ 2004; Desai et al.\ 2004; Gill et al.\ 2004a; Diemand et al.\
2004c; Gao et al.\ 2004b; Weller et al.\ 2004; Reed et al.\ 2004;
Nagai \& Kravtsov 2004). In particular, $\Lambda$CDM haloes appear to
have almost identical substructure, consistent with the results of
paper II, and the intrinsic variation in the
cumulative distributions from one halo to another are similar to
those reported here.

\section{Comparison between Numerical and Semi-analytic Results}\label{sec:4}

\subsection{Cumulative distributions}\label{subsec:4.1}

We now turn to the comparison between numerical and semi-analytic
results. We will consider results for the dense inner regions of the
halo and the lower-density outer regions separately, since numerical
effects may affect the former to a greater degree, as discussed in
section \ref{subsec:4.3} below. The right-hand panel of Fig.\
\ref{fig:2} shows the cumulative mass function, for all substructure
between 0.5 and $1.0\,r_{\rm vir,m}$ from the centre or each halo. Over
this range of radii, our model reproduces the numerical results almost
exactly, both in normalisation and in scatter. For massive haloes,
the cumulative distribution in the semi-analytic haloes is very
similar to those in the high-resolution simulations. All three
simulations lie below our average value, but the offset is a small
($\sim 20$ percent, or about equal to the halo-to-halo scatter, on
average), so it may not be significant. There are several effects
such dynamical age that could explain this offset, but we do not
expect our prescription for mass loss to be accurate to much better
than 10--20 percent in any case, as discussed in paper I.

\begin{figure}
 \centerline{\psfig{figure=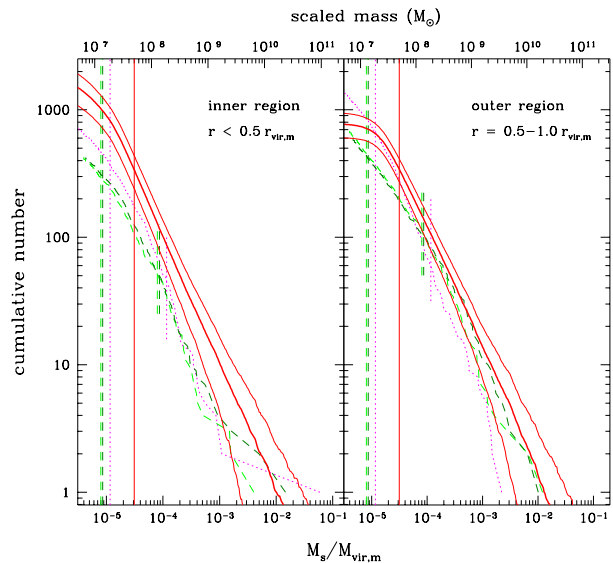,width=1.0\linewidth,clip=,angle=0}}
 \caption[]{The cumulative mass function predicted by the
semi-analytic model, in the inner (left-hand panel) and outer
(right-hand panel) parts of the halo. The thick lines show the average
result for a hundred SCDM merger trees, for model B at $z = 0$. The
thin solid lines show the 1-$\sigma$ variance for this set. The thin
lines are the normalised cumulative mass functions measured in the
three highest-resolution simulations (dashed lines -- Virgo IIa and
IIb; dotted lines -- Coma). The vertical dotted and dashed lines
indicate the 32-particle and 320-particle resolution limits of the
numerical results. The solid vertical line indicates the resolution
limit of the semi-analytic trees. }\label{fig:2}
\end{figure}

\begin{figure}
 \centerline{\psfig{figure=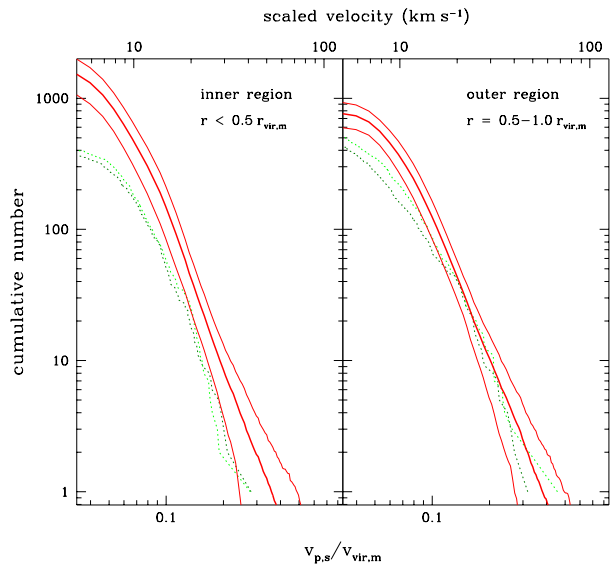,width=1.0\linewidth,clip=,angle=0}}
 \caption[]{The cumulative peak circular velocity functions predicted
by the semi-analytic model. The left-hand panel shows results for
haloes within half the virial radius; the right-hand panel shows
results for haloes between 0.5 and $1\,r_{\rm vir,m}$. The thick lines
are the average result and $\pm$ 1-$\sigma$ contours for a hundred
SCDM merger trees, for model B at $z = 0$. The thin lines are the
normalised cumulative velocity functions measured in the Virgo
IIa and IIb simulations. }\label{fig:3}
\end{figure}

At smaller masses ($M_{\rm s} < 10^{-4} M_{\rm vir,m}$), the
semi-analytic model predicts 30--40 percent more substructure above a
given mass threshold. It seems likely that at least some of this
offset is due to numerical effects such as relaxation, since here we
are below the limit of a few hundred particles where the relaxation
time becomes shorter than the Hubble time (Diemand et al.\ 2004a).
Overall we conclude that in the outer part of the halo, where the
properties of substructure are most robustly determined in the
simulations, the two sets of results are in acceptable agreement.

On the other hand, in the region interior to $0.5\,r_{\rm vir,m}$
(left-hand panel), the semi-analytic mass function predicts roughly
2.5 times more substructure above a given mass threshold than is seen
in the numerical simulations. In terms of the halo-to-halo scatter,
all three numerical mass functions lie 2$\sigma$ below the average
value in the semi-analytic trees. The cumulative velocity functions
(Fig.\ \ref{fig:3}) show a similar pattern. This suggests that the
two methods disagree significantly about how quickly substructure is
stripped or destroyed in the central regions of a
halo. Unfortunately, it is not clear which result is more
accurate. As seen in paper II, central subhaloes are
generally older and they will have experienced more mass loss and
tidal heating on average, having orbited many times in a strong and
changing potential. Since many of these central systems will be
heavily stripped, the semi-analytic predictions about their residual
bound mass be less accurate than for younger systems. On the other
hand, the simulations will also be less accurate for old systems and
at small radii, due to the cumulative effects of relaxation and
artificial heating. Moreover, it is harder for group finders to
identify substructure correctly in dense regions (Gill et al.\
2004a), and the subhalo masses and velocities they determine in these
regions can be biased by the background density. Thus it may be that
semi-analytic predictions for substructure are in fact {\it more}
accurate than simulations in the centres of halos (say within
$0.3\,r_{\rm vir,m}$). We will discuss this further in section
\ref{subsec:4.3}.

\subsection{Individual subhaloes and the role of softening}\label{subsec:4.2}

\begin{figure}
 \centerline{\psfig{figure=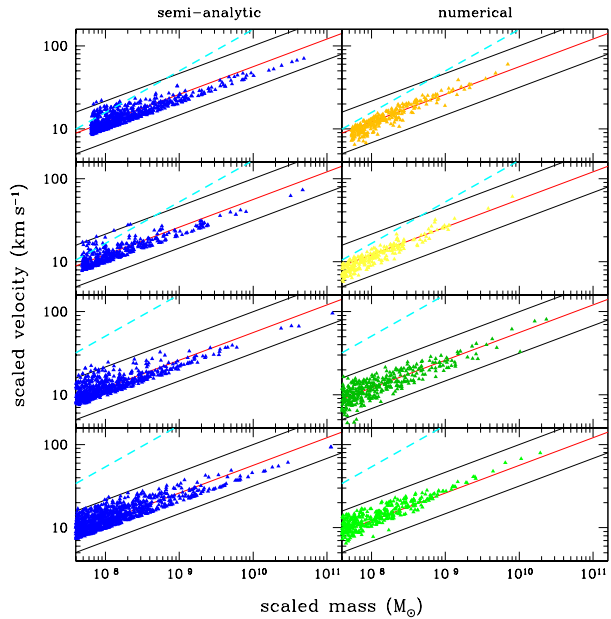,width=1.0\linewidth,clip=,angle=0}}
 \caption[]{The distribution of subhaloes as a function of their mass
and of their peak velocity, in the semi-analytic model A (left-hand
plots) and the simulations (right-hand plots). The numerical values
have been scaled to the mass and circular velocity of the main halo
(see text). The dashed lines indicate the regions of the plot excluded
by softening. }\label{fig:4}
\end{figure}

We can also compare the properties of individual haloes directly.
Fig.\ \ref{fig:4} shows a comparison of the semi-analytic subhaloes
(left-hand plots) and the numerical subhaloes (right-hand plots), in
terms of their mass and their peak circular velocity. The numerical
results, from top to bottom, are from the Milky Way, Andromeda, Virgo
IIb, and Virgo IIa haloes. The masses and velocities in the
simulations have all been rescaled to the mass and velocity of the
parent halo in the semi-analytic model, as explained in section
\ref{subsec:3.1}, and in each pair of panels we have only plotted
systems above the mass-resolution limit of the numerical data in the
right-hand panel.

Overall, the distributions seem remarkably similar. Comparing them in
detail, however, we note some minor differences. The semi-analytic
model predicts the existence of low-mass, high-density 
(high-$v_{\rm p}$) systems, for instance, which are not seen in the
simulations. This is partly because the forces in the simulation are
softened over a finite length $r_{\rm s}$, such that the potential
generated by a set of particles of mass $M$ is limited to 
$\simeq GM/r_{\rm s}$, placing a corresponding limit on $v_{\rm p}$. 
The dashed lines in each of the left-hand panels indicate the locus of
this limit, for the values of $r_{\rm s}$ listed in table 1. 
As expected, none of the numerical subhaloes lie
above this line, whereas we might expect some to in the low-resolution
simulations (upper two panels). The high-resolution simulations
(lower two panels) fall well short of this limit, although they may be
still subject to relaxation and other effects.

We also see that the numerical distributions generally extend to lower
circular velocities at a given mass than the semi-analytic
distributions. This may be partly due to softening, but another
explanation is shot noise in the particle distribution for these
systems. Low mass haloes will have few particles interior to 
$r_{\rm p}$, so subtracting a single particle from a halo can reduce 
its peak
velocity substantially. This may explain the greater scatter in the
lower left-hand corner of each of the numerical distributions.

\begin{figure}
 \centerline{\psfig{figure=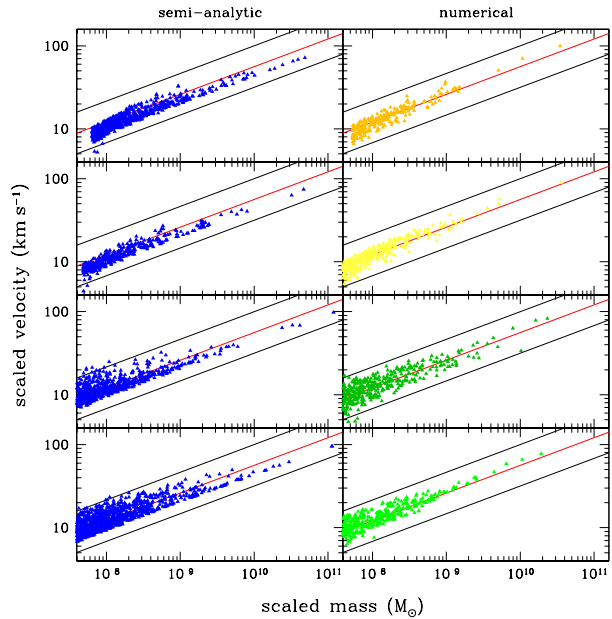,width=1.0\linewidth,clip=,angle=0}}
 \caption[]{As Fig.\ \ref{fig:4}, but including some of the effects
of force softening and mass resolution in the semi-analytic results
(left-hand panels).}
\label{fig:5}
\end{figure}

We can model the effect of softening explicitly by assuming that the
circular velocity is determined by the radial force, 
$v_c = r\dot F_r$, and using the force softening to reduce $v_c$ 
accordingly. The
forces in these simulations were spline softened, that is the
potential generated by each particle was calculated as 
$\Phi = \Phi_{\rm s}\,(r/r_{\rm s})$, where $\Phi_{\rm s}$, the 
softened potential, is a polynomial $P_1(r/r_{\rm s})$ for 
$0 \le r \le r_{\rm s}$, a polynomial $P_2(r/r_{\rm s})$ for 
$r_{\rm s} \le r \le 2r_{\rm s}$, 
and equal to the Newtonian potential beyond this (where $r_{\rm s}$ 
is the softening length of the simulation). We can account for
this by reducing the radial force accordingly; this reduces the peak
velocity when $r_p$ is close to the softening length $r_s$. To
simulate shot noise, we can assume that the number of particles within
$r_p$ varies randomly by $\sqrt N$, thereby introducing a scatter into
haloes where $M(<r_p)$ is close to $m_p$, the particle mass. Fig.\
\ref{fig:5} shows the distribution of velocities and masses, with
both shot noise and softening taken into account. We see that our
modified distributions are now very close to those found in the
simulations, particularly at low resolution.

\begin{figure}
 \centerline{\psfig{figure=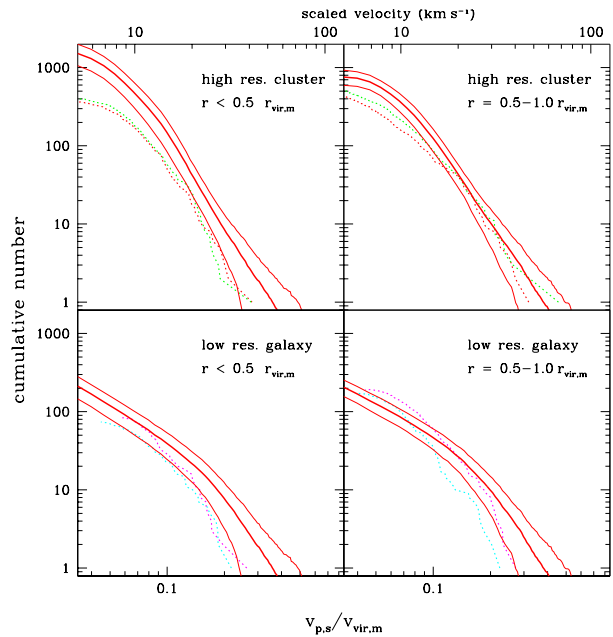,width=1.0\linewidth,clip=,angle=0}}
 \caption[]{Average cumulative peak circular velocity functions
predicted by the semi-analytic model, including the effects of
softening. Merger trees and line styles are as in Fig.\ \ref{fig:3}.
The softening length in the upper panels, $r_{\rm s} = 0.0005\,r_{\rm
vir,m}$, is comparable to that in the Virgo IIa and IIb simulations
(dotted lines), while the lower panels show similar results using a
softening length of $0.005\,r_{\rm vir,m}$, along with the normalised
Local Group distributions (dotted lines). }
\label{fig:6}
\end{figure}

Finally, we can re-examine the cumulative velocity function with
softening taken into account. Fig.\ \ref{fig:6} shows the cumulative
(peak circular) velocity function for subhaloes, with line styles as
before. The semi-analytic results have been softened as in Fig.\
\ref{fig:5}, with a softening length corresponding to that used in
the simulations shown. The upper panels are for the higher-resolution
Virgo II simulations, in which the softening length was 
$r_{\rm s} = 0.0005\,r_{\rm vir,m}$ (or 170\,pc in our fiducial system), 
while the
lower are for the Local Group simulations, in which the softening
length was roughly $r_{\rm s} = 0.005\,r_{\rm vir,m}$ (or 1.7\,kpc in
our fiducial system). As before, the semi-analytic predictions match
the simulations reasonably well in the outer parts of the halo, but
predict 2--3 times more substructure above a given velocity threshold
in the inner parts. Comparing the upper and lower panels, we see that
softening alone may account for most of the difference between the
high-resolution and low-resolution numerical results in the amplitude
of the cumulative velocity function below $v_{\rm s}/v_{\rm vir,m}
\sim 0.15$. The match between the softened semi-analytic predictions
and the simulations is still not exact, however (e.g.\ the
disagreement in the amplitude of the mass function at 
$r < 0.5\,r_{\rm vir,m}$), suggesting there may be other resolution 
effects we have not considered.

Indeed, there are several well know sources of artificial heating in
$N$-body simulations that we have not accounted for so far. Internal
relaxation will reduce the mass, circular velocity and potential of
each subhalo artificially, on a timescale roughly proportional to the
number of particles. For systems resolved with fewer than $\sim 300$
particles, this timescale is shorter than the Hubble time, as
mentioned previously, so only the youngest objects will be unaffected
by relaxation. The `graininess' of the background potential will also
heat systems artificially, particularly at early times when the main
halo is poorly resolved. These effects have been studied extensively
in the literature in the context of the `overmerging problem', as
discussed in the next section.

\subsection{Spatial distributions and the evidence for overmerging}\label{subsec:4.3}

\subsubsection{Radial distributions compared}
The results of section \ref{subsec:4.1} suggest that the simulations
may underestimate the amount of substructure in the central regions of
haloes. In early simulations, the dissolution of substructure within
haloes, referred to as `overmerging', rendered simulated systems
almost completely smooth (see van Kampen 1995; Moore, Katz, \& Lake
1996; or Klypin et al.\ 1999a for discussions of the
problem). Overmerging is known to occur to some degree even in
high-resolution simulations (Ghigna et al.\ 2000), and should be
strongest in the old, dense central regions of haloes (Diemand et
al.\ 2004a). The results of section \ref{subsec:4.1} suggest that it
could still be important over a fairly large range of radii.

\begin{figure}
 \centerline{\psfig{figure=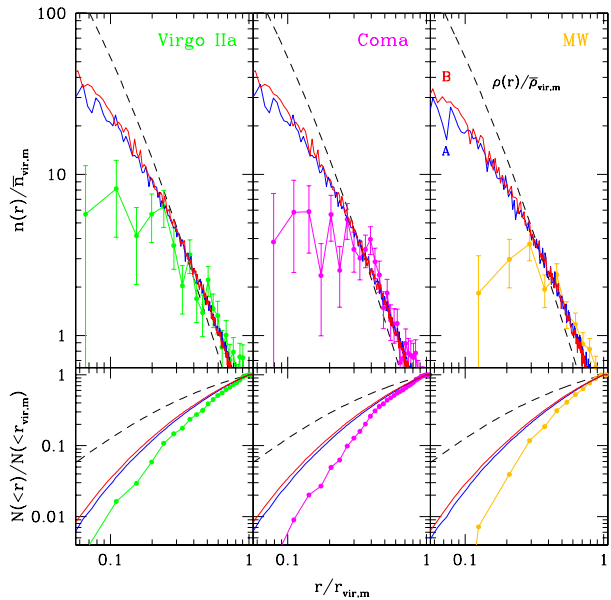,width=1.0\linewidth,clip=,angle=0}}
 \caption[]{Top panels: The number density of subhaloes in three
simulations (connected points with error bars), and in the
semi-analytic haloes (upper solid lines), for models A and B. To
avoid incompleteness, the semi-analytic results include only systems
with masses in excess of $5 \times 10^{7} M_{\odot}$ (left and middle
panels) or $10^{8} M_{\odot}$ (right panel), and the numerical
results have been restricted to the same relative mass range. In each
case the density is relative to the mean number density within the
virial radius. The dashed line shows the density profile of the main
halo, normalised to the mean density within the virial radius. Bottom
panels: The cumulative number of subhaloes vs.\ radius, normalised to
the number within the virial radius, for the same mass cuts as in the
top panel. The dashed lines show the mass of the main halo interior
to a given radius, normalised to the mass within the virial radius.}
\label{fig:7}
\end{figure}

We can quantify the effects of overmerging by comparing the radial
distribution of substructure in our model and in the simulations. 
The top three panels of Fig.\ \ref{fig:7} show the local density 
of subhaloes at a given radius, relative to the mean density within 
the virial radius, 
$n(r)/\overline{n}_{\rm vir,m} = (N(<r)/dV(r))/(N_{\rm vir}/V_{\rm vir})$. 
The connected points with error bars show the
results in three simulations, and the upper solid lines show the
predictions of semi-analytic models A and B. We saw in 
paper II that the radial distribution of subhaloes is
biased by incompleteness if we go below the mass resolution limit of
the merger tree. To avoid this bias, the semi-analytic results shown
in the left and middle panels include only systems with masses in
excess of $5 \times 10^{7} M_{\odot}$, while the numerical results
are limited to an equivalent relative mass range, 
$M_{\rm s}/M_{\rm vir,m} > 3\times 10^{-5}$. The resolution limit of 
the `MW' simulation is actually worse than this, so in the right-hand 
panel we cut both the numerical and the semi-analytic results at 
$10^{8} M_{\odot}$. The dashed line shows a Moore density profile of
concentration $c_{\rm M} = 5.4$ (roughly appropriate for galaxy or
cluster mass haloes), also normalised to the mean density within the
virial radius. We note that similar numerical results have been
presented recently by several authors (Gill et al.\ 2004a; 
Diemand et al.\ 2004c; Gao et a.\ 2004; Reed et al.\ 2004; 
Nagai \& Kravtsov 2004).

The local density profile has the disadvantage of being quite noisy in
the central regions of the halo, and its overall appearance depends
partly on the choice of radial bins. In the bottom panels, we
therefore show the cumulative number of subhaloes within a given
fraction of the virial radius, normalised to the total number within
the virial radius, since this quantity is monotonic and requires no
binning. The mass cuts are the same as in the top panel, and the dashed
lines show the mass of the main halo interior to a given radius,
normalised to the mass within the virial radius.

Both numerical and semi-analytic models agree that subhaloes are
antibiased with respect to the underlying density distribution, and
both agree on the distribution in the outer parts of the halo, at $r
> 0.3\,r_{\rm vir,m}$. In the central region, however, the
semi-analytic model predicts a substantial excess of subhaloes
compared to the simulations -- $n(r)/\overline{n}_{\rm vir} \sim 20$
at $0.1\,r_{\rm vir,m}$ and $\sim 10$ at $0.2\,r_{\rm vir,m}$,
whereas in Virgo IIa the values are $\sim 7$ and $\sim 5$
respectively. As an indication that the semi-analytic result is
robust, we see that the excess depends only weakly on the disruption
criterion used (the upper and lower semi-analytic curves correspond to
models B and A respectively). On the other hand, with increasing
resolution (three panels, right to left) the numerical distributions
gradually become more concentrated, approaching the semi-analytic
results in the highest-resolution case. Thus it seems likely either
that overmerging is still important in the inner regions of the
simulated haloes, or that the group finders used to generate the
numerical datasets have missed substructure in the central regions. 
We will discuss this further in section \ref{subsec:4.5}.

Overmerging at the level we are suggesting should also reduce the
amplitude of the cumulative mass function within the virial radius,
but the overall effect will be small, because even in our
semi-analytic models, relatively few subhaloes at found at small
radii. Since the semi-analytic model predicts that only 25--30 percent
of all haloes within the virial radius are at radii of 
0.2--$0.3\,r_{\rm vir,m}$ or less, the change in the amplitude of the 
mass function would only be 25--30, even if overmerging destroyed all
objects in these regions. This may explain why simulations have
previously shown good convergence in the cumulative distributions 
of subhaloes within the virial radius as a function of resolution 
(e.g.\ Springel et al.\ 2001; Diemand et al.\ 2004c; Gao et al.\
2000b). These distributions are dominated by subhaloes relatively 
far from the centre of the potential, which are less influenced by
numerical effects, and thus they will not be sensitive to central
overmerging.

On the other hand, it seems more surprising that convergence studies
have seen no major change in the radial distribution of substructure
(Diemand et al.\ 2004c; Nagai \& Kravtsov 2004). This may be partly
due to the obscuring effects of halo-to-halo scatter, halo
concentration or binning, which make it difficult to identify
statistically significant differences between two density
distributions. It may also be that the convergence in the radial
distribution of substructure is very slow; we will discuss this
further in section \ref{subsec:4.5}.

\subsubsection{Results for variant models}
Overmerging of the magnitude suggested by these results would have
important implications in many astrophysical situations, notably the
interpretation of strong lensing observations and direct detection
experiments. Thus, it is interesting to consider how strongly these
results could be affected by uncertainties in the semi-analytic
modelling. We have compared number density profiles for the variants
of the model considered in paper II with our fiducial
results. While the profiles change in predictable ways (e.g.\ more
dynamical friction or less mass-loss produces more central
substructure over a given mass threshold), the variation is generally
comparable to the difference between models `A' and `B' shown in 
Fig.\ \ref{fig:7}.

\begin{figure}
 \centerline{\psfig{figure=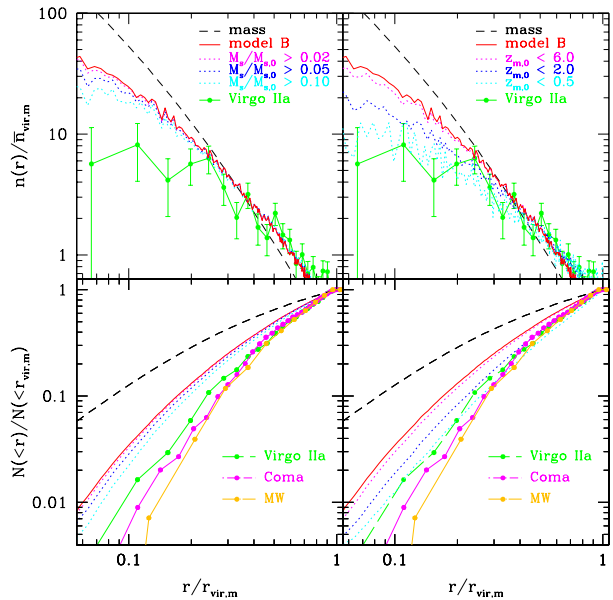,width=1.0\linewidth,clip=,angle=0}}
 \caption[]{As Fig.\ \ref{fig:7}, but for various cuts in subhalo
properties. The left-hand panels show the results of ignoring all
subhaloes stripped beyond some fraction of their original mass (dotted
lines); the right-hand panels show the results of ignoring all
systems the formed before a given epoch (dotted lines). }
\label{fig:8}
\end{figure}

On the other hand, it might be that our analytic mass-loss model
systematically underestimates mass loss in systems that have been
heavily stripped. To get a sense of how large an effect is required to
reproduce the numerical results, we have calculated number density
profiles excluding systems that retain only 2 percent, 5 percent, 
or 10 percent of their original mass. These are shown
in the left-hand panels of Fig.\ \ref{fig:8} (dotted lines), along
with the profiles from the three simulations (solid lines with points
-- note the mass resolution limit for the Milky Way results is
higher) and the fiducial results for model B (uppermost solid line).
We see that even if we treat as disrupted all systems that have lost
90 percent of their mass, we still produce more central substructure
than the highest-resolution simulation, albeit only by a factor of 2
or so. The results of Hayashi et al.\ suggest that bound cores can
survive in systems that have lost 99 percent of their mass or more, so
it seems unlikely that our mass-loss predictions are incorrect to a
degree sufficient to resolve the discrepancy with the numerical
results.

We can also get a feel for the plausibility of substantial numerical
overmerging by considering {\it how long} subhaloes have orbited
within the main system. The right-hand panels of Fig.\ \ref{fig:8}
show number density profiles excluding the oldest subhaloes, those
that first formed at redshifts of more than 6.0, 2.0 or 0.5 (dotted
lines). As expected from the results of paper II,
substructure is stratified with respect to its age, so the central
substructure we predict in excess of that found in the simulations is
mainly old -- almost all of the central systems formed before a
redshift or 0.5, when the universe was roughly half its present age,
and most formed before $z = 2$, when the universe was less than 20
percent of its present age. This material would have undergone many
orbits in the dense central regions of the main system or its
progenitors, so it seems very plausible that artificial numerical
heating could have caused it to disrupt prematurely. We will
reformulate this argument more precisely in the next section. Finally,
we note that while the radial distributions of substructure do vary
systematically from one halo to another if we bin haloes by their
formation epoch, as in paper II, the variation is
generally small (comparable to the difference between models A and B).

\subsection{Subhalo kinematics}
\label{subsec:4.4}

\begin{figure*}
 \centerline{\psfig{figure=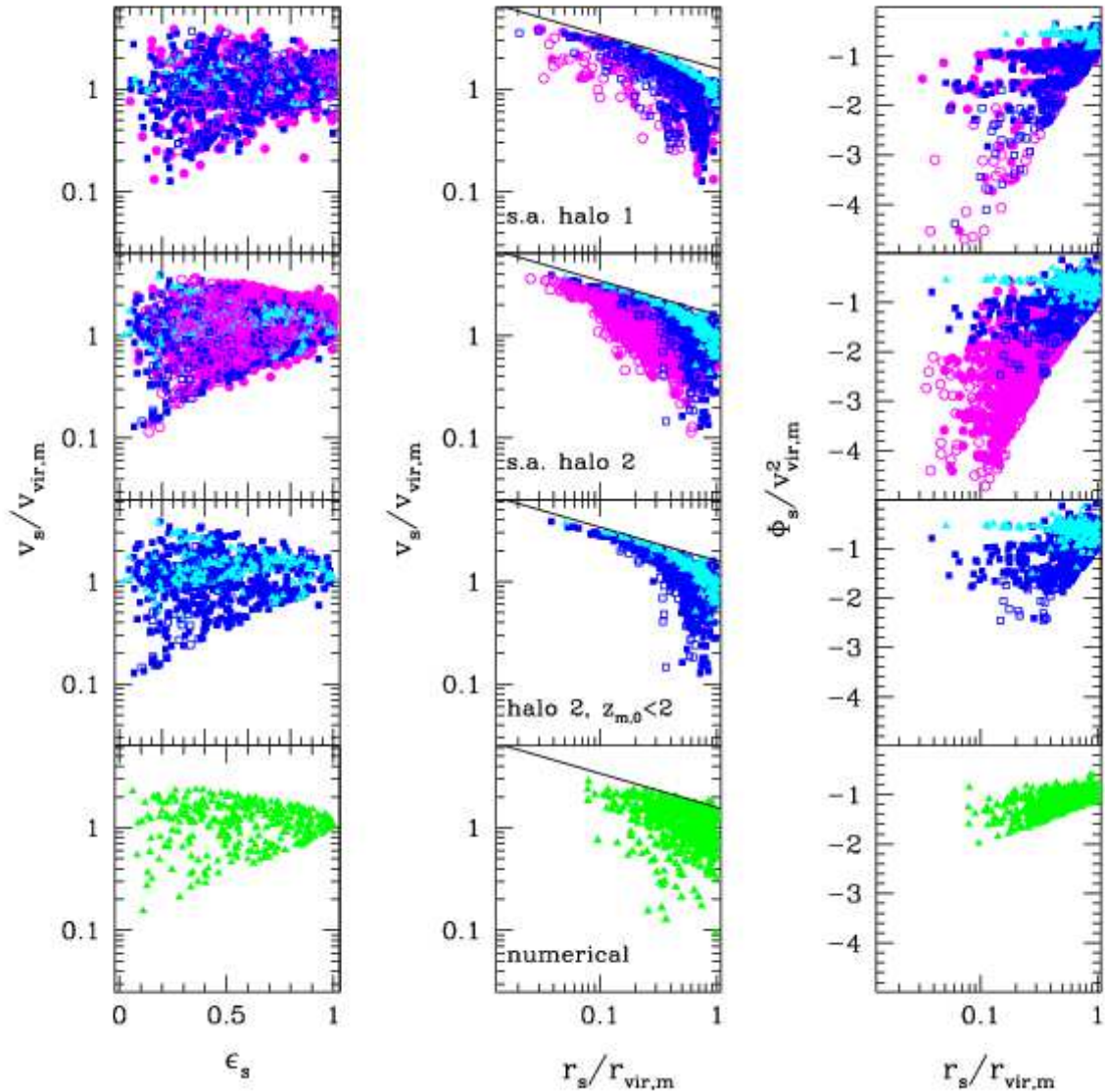,width=0.9\hsize,clip=,angle=0}}
 \caption[]{Kinematics and orbital parameters for subhaloes in two
different semi-analytic haloes (model B; first and second panels in each
column), subhaloes from the second of these that formed after $z = 2$
(third panel), and subhaloes in the Virgo IIa simulations (bottom
panels). The left-hand column shows velocity vs.\ circularity; the
middle column shows velocity vs.\ position, and the right-hand
column shows orbital energy vs.\ position. For the semi-analytic results,
open symbols represent systems that have lost more than 90 percent of
their original mass, while the symbol shape indicates formation epoch
(triangles: $z_{\rm m,0} < 0.5$; squares $z_{\rm m,0} = 0.5$--2.0;
circles $z_{\rm m,0} > 2.0$). In each case, all subhaloes within the
virial radius and over a mass limit of $10^{-5} M_{\rm vir,m}$ are
included.}
\label{fig:9}
\end{figure*}

If overmerging is important, it will also affect the distributions of
other subhalo properties. Fig.\ \ref{fig:9} compares the kinematics
and dynamical state of subhaloes in the semi-analytic and numerical
models. The top two panels in each column show subhaloes from two
different semi-analytic haloes; the third panel shows all subhaloes
from the second of these that formed at $z_{\rm m,0} < 2$, and the
bottom panel shows subhaloes from the Virgo IIa simulations. For the
semi-analytic results, open symbols represent systems that have lost
more than 90 percent of their original mass, while the symbol shape
indicates formation epoch (triangles: $z_{\rm m,0} < 0.5$; squares
$z_{\rm m,0} = 0.5$--2.0; circles $z_{\rm m,0} > 2.0$). In each case,
all subhaloes within the virial radius and over a mass limit of
$10^{-5} M_{\rm vir,m}$ are included.

The left-hand column shows velocity versus orbital circularity. In
paper I, we discussed the initial and final circularity distributions
in our model. As Fig.\ \ref{fig:9} shows, the final circularity and
velocity distributions for a semi-analytic system and the Virgo IIa
subhaloes are very similar. Given that the orbital properties of
subhaloes in the semi-analytic model are the result of a complex
superposition of several effects, including the initial energy and
angular momentum distributions, dynamical friction, selective
disruption and the growth of the main halo, this agreement is very
encouraging.

The middle column shows velocity versus position. Both in the
semi-analytic and in the numerical results, the distribution is
bounded by the same well-defined upper limit at any given radius. The
line indicates that this boundary is roughly 
$v_{\rm max}(r) = v_{\rm vir,m}(r/r_{\rm vir, m})^{-1/3}$ down to 
$r/r_{\rm vir, m} = 0.1$. The
semi-analytic model clearly predicts more substructure in the central
regions, and thus a higher central velocity dispersion for the
subhaloes as a group.

Finally, the right-hand column shows orbital energy versus
position. The overall distributions are very different, the
semi-analytic model predicting many more very strongly bound
subhaloes. Most of these systems are very old, however, and disappear
if we restrict the sample to systems that formed after $z = 2$ (third
panel from the top). Thus we see the same effect discussed in the
previous section, namely that the older systems predicted in the
semi-analytic models are absent in the numerical results.

\subsection{Comparison with semi-analytic results: summary}
\label{subsec:4.5}

In summary, in this section we have used a set of high-resolution
simulations to estimate the average properties of halo substructure,
as well as the intrinsic scatter from one halo to the next, and the
variation with halo mass or concentration. Comparing these
simulations with the predictions of our semi-analytic model, we find
that while there is an overall similarity in the results, the level
of agreement depends on the location, mass and age of the subhaloes.

\subsubsection{The outer halo}

In the case of intermediate or high-mass subhaloes in the outer
regions of the halo, for which the numerical results are expected to
be most reliable, the agreement between the two methods is excellent;
the cumulative mass and velocity distributions of the three
high-resolution simulations all lie within 1--2 times the halo-to-halo
scatter of the average value predicted by the semi-analytic model, and
the overall difference between the average semi-analytic and numerical
results is less than 20 percent.

Assuming this offset is significant, there are several effects that
could introduce systematics at this level. Possible effects in the
semi-analytic model include the various approximations in the
dynamical component of the model, harassment between subhaloes (cf.\
paper II), or the preferential selection of haloes
with older or younger formation epochs. On the latter point, we note
that the simulations discussed here generally selected relaxed
systems from larger volumes to study at high resolution; thus they do
not constitute an unbiased sample of the dark matter haloes in a given
mass range. The Virgo simulations, for instance, were of a cluster
that had acquired 80 percent of its final mass by a redshift of 0.75,
which is unusual for an object in this mass range (G98, Fig.\ 3). We
can see from paper II, Fig.\ 14 that if we were to select out the
oldest merger trees from our sets of semi-analytic haloes, we would
obtain an even closer match to the simulations.

Possible effects in the numerical results include softening, shot
noise, or problems with the group finder, all of which change the
interpretation of the results from a single output of the simulation,
as well as some more serious problems, notably two-body relaxation,
which actually modify the dynamics of simulated systems. The
increased offset between the semi-analytic and numerical results at
small masses may indicate the greater importance of these effects in
poorly-resolved systems. Overall, however, we conclude that for
intermediate or high-mass subhaloes in the outer regions of the halo,
the two methods are consistent with each other to good accuracy. We
note that this agreement is achieved without adjusting any free
parameters -- the parameters in the semi-analytic model have all been
fixed previously by other considerations, as discussed in paper I.

\subsubsection{The inner halo}

On the other hand, in the inner regions of the halo, where numerical
effects may be stronger, the semi-analytic model predicts
substantially more substructure than the simulations. Some of
this difference can be attributed to the same numerical effects
mentioned above, but the net effect is that the central regions of
haloes appear to suffer from a fair amount of overmerging.

The possibility of central overmerging and the spatial distribution of
substructure in numerical simulations have been investigated by a
number of authors in more recent simulations (Gill et al.\ 2004a; 
Diemand et al.\ 2004c; Gao et a.\ 2004; Reed et al.\ 2004; Nagai \&
Kravtsov 2004). On the one hand, convergence studies using a given
code and group finder find little or no evidence for a rapid increase
in the amount of central substructure as the resolution increases
(e.g.\ Diemand et al.\ 2004c; Nagai \& Kravtsov 2004). This might seem
a conclusive argument against overmerging, since increased resolution
is the only way of testing for this possibility definitively.

On the other hand, detailed studies of the convergence of a different
feature of halo structure, the slope of the central density profile,
have found that the size of the region flattened by resolution
effects such as relaxation decreases very slowly as the number of
particles in the halo $N$ increases, scaling as $N^{-0.2}$--$N^{-0.3}$
(Diemand et al.\ 2004a). This sort of scaling would be expected if the
size of the unresolved region depends the mean inter-particle
separation, for instance, or on the maximum density allowed by
softening.

Resolving substructure in a dense background environment represents a
similar, but harder, numerical problem. Not only does finite
resolution limit the density of structures that can be resolved, but
subhaloes are intrinsically less stable than a central cusp, since
they are subject to much stronger tidal forces. Thus we we should not
expect to be able to resolve substructure at densities or on spatial
scales where the central cusp of the main halo is flattened by
relaxation, and in general the size of the region where substructure
is artificially erased by relaxation should decrease no faster than
$N^{-1/3}$ as the number of particles $N$ increases.

\subsubsection{How large is the unresolved region?}

We can apply this argument more specifically to the numerical results
presented here. The Milky Way and Virgo IIa simulations agree in the
distribution of substructure beyond $\sim 0.3\,r_{\rm vir}$, but
disagree within this radius. Given that they differ in $N$ by a factor
of 5, we estimate that the Virgo IIa results are reliable to $\sim
0.17\,r_{\rm vir}$. Diemand et al.\ (2004b, Fig.\ 7) present results
for 4 galaxy haloes simulated with 1--4 million particles each. At
$0.17\,r_{\rm vir}$ they find $n(r)/\overline{n}_{\rm vir}\simeq$ 10,
consistent with the semi-analytic predictions and the Virgo IIa
results, so this may well be the radius at which the numerical
results have converged.

Diemand et al.\ (2004b, Fig.\ 2) also present even higher resolution
results, for a cluster halo resolved with up to 14 million
particles. By the same scaling argument, we would expect these to be
reliable down to $0.11\,r_{\rm vir}$, but in fact they still see
substantially less substructure at this radius than is predicted by
our model ($n(r)/\overline{n}_{\rm vir}\simeq$ 4--5, versus 15--20 in
our model). On the other hand, their results for this cluster differ
by a factor of 2 with the results for their four galaxy haloes (for
which $n(r)/\overline{n}_{\rm vir}\simeq$ 10 at this radius). Thus it
is unclear whether the shallower density profile is characteristic of
the cluster mass scale as opposed to the galaxy mass scale, whether it
is simply due to intrinsic halo-to-halo variation, or whether the
numerical convergence is even slower than $N^{-1/3}$.

There are indications of the first of these possibilities in the
results of De Lucia et al.\ (2004, Fig.\ 6) and Gao et al.\ (2004b,
Fig.\ 11), who find that subhaloes follow a more centrally
concentrated distribution in galaxy haloes than in cluster halos. With
regards to the second possibility, Gill et al.\ (2004b, Fig.\ 7) show
that that the scatter in the number density profile can be a factor of
2 or more in amplitude, although all 8 of their haloes have steep inner 
slopes in the radial distribution of substructure. In any case, if the 
third explanation were correct and the convergence rate scaled as
$N^{-0.2}$, we would expect convergent results only beyond 
$0.17\,r_{\rm vir}$, \ so this could also explain the discrepancy.

Finally, we note that the algorithm used to locate and define
substructure may have a large effect in and of itself. Weller et al.\
(2004), for instance, find quite different bound mass estimates for
subhaloes depending on the criterion used for associating particles
with substructure, while Gill et al.\ (2004a), obtain substantially 
different results for the radial distributions of subhaloes by 
`tracking' halo particles from one step to the next. In particular, 
they find that in all 8 of their haloes, the radial density of subhaloes 
identified by `tracking' continues to rise down to the smallest radii 
they consider, $r \simeq 0.07\,r_{\rm vir}$ (cf.\ their Fig.\ 7).
With this technique, 10 percent of the substructure they identify is
located within the inner 0.1--$0.2\,r_{\rm vir}$, as in our semi-analytic
model, whereas for a group finder using only information from a single 
timestep, they find the cumulative distribution reaches 10 percent at
$\sim 0.3\,r_{\rm vir}$, as in the simulations considered here. Analysing
the same halo with different group finders, they find that normalised 
density at $0.1\,r_{\rm vir}$ varies by a factor of 4.

\subsubsection{Relaxation times for central subhaloes}

There is a second argument that suggests that simulations may still be
missing substructure in their central regions. We expect strong
correlations between the age of subhaloes and their location within
the main system (cf. paper II). Central subhaloes are
systematically older, and if we artificially remove the oldest
subhaloes from our semi-analytic results, we achieve a much better
match to the numerical results (cf.\ Figs.\ \ref{fig:8} and
\ref{fig:9}). Central subhaloes typically formed at or before $z = 2$; 
thus they were originally $3^3 = 27$ times denser than present-day
systems of the same mass, and have spent roughly 5 orbits or 11 Gyr in
the main system (paper II, Fig.\ 8), losing 75 percent of their
original mass in the process (paper II, Fig.\ 9).

On the one hand, the high-resolution simulations of H03 and
Kazantzidis et al.\ (2004) indicate that low-density systems resolved
with $\sim 10^4$ particles can easily survive this degree of mass loss
without disintegrating. On the other hand, when systems of the density
corresponding to $z_{\rm m,0} = 2$ are resolved with fewer than 5000
particles, their relaxation time is less than the Hubble time
(Diemand et al.\ 2004). For systems of $\sim 32$ particles, the
nominal resolution limit of the numerical data sets, the relaxation
time at this density is 100-200 Myr. While it is not clear how exactly
quickly relaxation leads to the disruption of substructure, it seems
unlikely that 32 particle systems merging at $z = 2$ could survive for
the equivalent of 50 relaxation times without being completely
dissolved. This artificial disruption due to internal relaxation
would have little effect on the overall properties of substructure
averaged over the entire halo, since only a small fraction of all
subhaloes are this old. On the other hand it would quickly reduce the
central density of subhaloes, since the latter is dominated by small
objects that formed at early times.

The preferential disruption of old subhaloes close to the centre of
the main system has important implications for many of the
observational tests of halo substructure. In section \ref{sec:5}, we
will consider two examples, the detection of substructure in
multiply-lensed systems, and the direct detection of dark matter in
terrestrial experiments.

\section{Implications for Observational Tests of Substructure}\label{sec:5}

\subsection{Implications for lensing}\label{subsec:5.1}

Gravitational lensing, the deflection of light from a background
source by the gravitational potential of a foreground system, provides
at least two ways of quantifying the amount of dense substructure in
the halo of the lensing system. The dark matter around individual
galaxies systematically distorts the shapes of background galaxies
within some projected separation. In rich clusters, these distortions
can be averaged over a large number of galaxies at the same redshift,
yielding maps of the projected mass density within the cluster, and
statistical information about the subhaloes around each cluster member
(e.g.\ Natarajan, Kneib \& Smail 2002; Gavazzi et al.\ 2004). The
resulting mass maps cover a large fraction of the projected area of
the cluster at high spatial resolution, but they are somewhat
model-dependent. In particular, they require the presence of visible
galaxies to trace the dark matter substructure.

There is an alternate lensing method that promises to reveal completely
dark substructure in the haloes of galaxies. It consists of comparing
the amplification ratios of different components in multiply-lensed
systems with models of the mass distribution in the lens. For
particular lens geometries, discrepancies in the amplification ratios
may indicate a deviation from a smooth potential on the scale of the
image separation. This method for quantifying substructure received
much attention recently, with claims that the projected mass fraction
contained in substructure had been measured fairly reliably for
a set of systems (Dalal \& Kochanek 2001, 2002), and that lensing
statistics might be allow the power spectrum to be constrained
directly (Zentner \& Bullock 2003).

In the light of subsequent work, these results now seem less certain.
For many individual systems, various other effects including stellar 
microlensing
(Schechter \& Wambsganss 2002), scintillation, or biases in the lens
modelling (Evans \& Witt 2003) may be sufficient to explain the
anomalous flux ratios. An improved method uses observations at many
different wavelengths to eliminate the microlensing contribution,
taking advantage of the fact that microlensing and lensing by
substructure should have different effects on the broad-line and
narrow-line regions of lensed AGN, due to their different spatial
scales (Moustakas \& Metcalf 2003; Metcalf et al.\ 2004). On the other
hand, even this method cannot prove that the small-scale structure is
actually within a given halo, rather than simply being seen in
projection (e.g.\ Chen, Kravtsov, \& Keeton 2003; Metcalf 2004). In
the longer term, ultra-high resolution images from very long baseline
interferometry (Inoue \& Chiba 2003) or novel techniques with X-ray
telescopes (e.g.\ Yonehara, Umemura, \& Susa 2003) in space may 
produce more conclusive detections of halo substructure.

\begin{figure}
 \centerline{\psfig{figure=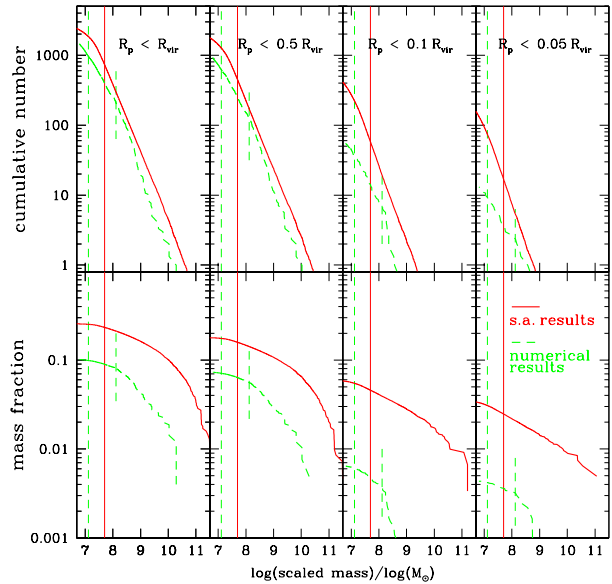,width=1.0\linewidth,clip=,angle=0}}
 \caption[]{(Top panel) Cumulative mass functions for subhaloes
within some projected radius $R_{\rm p}$, for model B (solid lines)
and Virgo IIa (dashed lines). (Bottom Panel) The fraction of the
projected mass within $R_{\rm p}$ contained in subhaloes of mass $M$
or larger. The numerical results are the average over three different
projections. Vertical lines indicate the resolution limit of them merger
tree (solid) and the 32 and 320-particle mass limits of the simulation.
}
\label{fig:10}
\end{figure}

Whatever the status of the problem observationally, it is not clear
that there is a robust theoretical prediction with which to compare
current observational results. Strong lensing probes the mass fraction in
relatively low-mass substructure ($10^5 M_{\odot}$--$10^7 M_{\odot}$), 
in the central few kiloparsecs of galaxy haloes. This is
well within the region where there is evidence for overmerging in the
simulations. We can estimate the importance of overmerging by
comparing the substructure in our semi-analytic models with the
substructure in the numerical simulations, as a function of projected
distance from the centre of the halo. Fig.\ \ref{fig:10} shows the
cumulative mass functions (upper panel) and cumulative mass fraction
(lower panel) for subhaloes within some projected radius $R_{\rm p}$,
for model B (solid lines) and Virgo IIa (dashed lines). (The numerical
results are the average over three orthogonal projections.)

Averaged over a large projected radius, the semi-analytic and
numerical results disagree by a factor of $\sim$2. At large masses,
some of this offset may be due to the dynamical age of the simulated
system, as discussed in paper II (cf.\ paper II, Fig.\ 15). 
At the low-mass end, relaxation or other resolution
effects may explain the offset, of it may be due to random,
halo-to-halo variation.

More worrying, however, is the offset between the numerical and
semi-analytic results at small projected radii. We noted in section
\ref{subsec:4.3} that in the central regions of the halo the density
of subhaloes is almost constant in the simulations, but continues to
rise in the semi-analytic model. As a result, the projected mass
fraction in substructure within the central 5 percent of virial radius
($\sim 15$\,kpc, or roughly the optical radius for a system like the
Milky Way) differs by an order of magnitude between the two
methods. This difference may be due to overmerging in the
simulations, as discussed in section \ref{subsec:4.3} and
\ref{subsec:4.5}, or it could reflect the limitations of the group
finder used to analyse the simulation, as discussed in Gill et al.\
(2004a).

It is premature to draw firm conclusions on the true projected mass
fraction from these results, for several reasons. First, the results
shown here are for SCDM, since the simulations used for comparison
with assumed this cosmology. The slightly reduced amplitude of the
cumulative mass function seen in LCDM haloes (see paper II) could 
affect the projected quantities to some
degree. Furthermore, while our dynamical model successfully reproduces
the evolution of systems during early stages of mass loss, it may be
less accurate for the subhaloes in the centre of the main system, many
of which have lost 90 percent of their mass or more (cf. paper II, Fig.\ 11). 
Based on Fig.\ \ref{fig:7}, a radical change in the 
properties of systems stripped to this degree might reduce the central 
density of subhaloes by a factor of 2, although this would still 
exceed the density found in the simulations.

Most importantly, however, the semi-analytic model predicts large
halo-to-halo scatter, which is correlated with the dynamical state of
haloes (cf. paper II, Fig.\ 15). Thus the comparison between
semi-analytic or numerical models and observed systems should
account for possible selection effects in the haloes considered. In
particular, selecting observed systems on the basis of a particular
galaxy morphology (e.g.\ ellipticals) may correspond to picking haloes
that are systematically more relaxed, and thus contain 2--3 times
less substructure than average. Finally, neither the
semi-analytic nor the numerical results presented here include a
galaxy in the potential of the main system, so {\it both} may
overestimate the amount of dark matter substructure in the central
regions to some degree.

Despite all these caveats, it is intriguing that our model predicts
projected central mass fractions roughly ten times larger than those
measured in high-resolution simulations, and that this in turn is
close to the value inferred both in early observational estimates
(Dalal \& Kochanek 2002) and in more recent detailed work (e.g.\
Metcalf et al.\ 2004). We will attempt to make more robust estimates
of the mass fraction in substructure and discuss the uncertainties 
in the semi-analytic modelling of central substructure in forthcoming
work.

\subsection{Implications for direct detection}\label{subsec:5.2}

Ultimately, the most convincing way to identify the dark matter
particle will be to detect it directly in a terrestrial
experiment. There has been a concerted effort for many years to search
for the mechanical effects of collisions between dark matter particles
and nuclei in calorimeters on Earth (see Pretzl (2002) for a recent
review). These experiments have gradually set more and more stringent
limits on the dark matter cross-section, without producing a
confirmed detection. Tentative evidence for a signal was announced by
the DAMA collaboration (cf.\ Bernabei et al.\ 2000, 2003), which
claimed to see an annual modulation in their event rate,
corresponding to the Earth's changing velocity with respect to the
distribution of dark matter in the halo, as it orbits around the
Sun. Work of comparable sensitivity by other experiments (e.g.\
ZEPLIN I -- Liubarsky et al.\ 2000; EDELWEISS -- Benoit et al.\ 2002;
CDMS -- Akerib et al.\ 2004) has failed to reproduce this result,
however, so it remains controversial (see Morgan, Green, \& Spooner
(2004) for a recent summary of the situation).

\begin{figure}
 \centerline{\psfig{figure=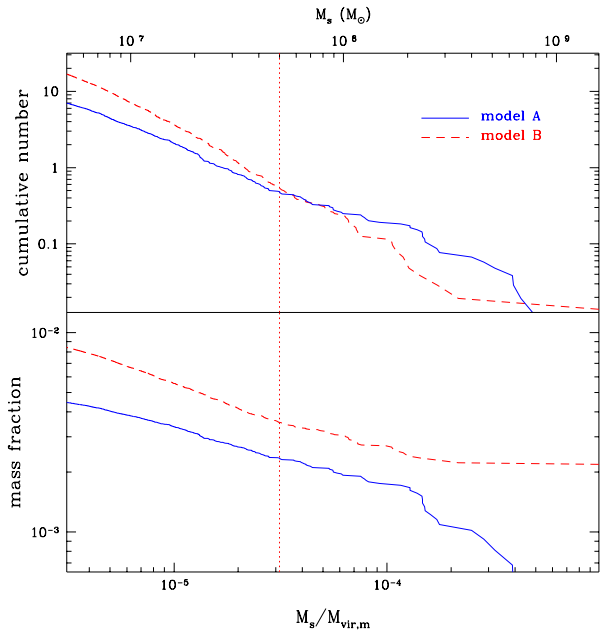,width=1.0\linewidth,clip=,angle=0}}
 \caption[]{Cumulative mass functions and contributions to the total
mass within the solar volume (6--10\,kpc). The dotted line is for
model A, the solid line is for model B.}
\label{fig:11}
\end{figure}

A crucial factor in interpreting the DAMA result is the regularity of
the phase-space distribution of dark matter particles in the solar
neighbourhood. Local substructure could introduce additional
modulations in the event rate, thereby reducing the sensitivity of
experiments looking for an annual signal (Morgan et al.\ 2004, and
references therein). Previous work on the local phase-space
distribution of dark matter based on numerical simulations 
(Helmi et al.\ 2003) found that substructure was rare in the solar
neighbourhood, so that the confusing effects of coherent streams
should not be a problem for direct detection experiments. Given the
evidence for overmerging in the central regions of simulated haloes
presented in section \ref{subsec:4.3}, however, this conclusion may
need to be revised.

We cannot easily compare our semi-analytic predictions with simulated
substructure around the position of the Sun, as there is essentially
no such substructure in the simulations -- the chance of finding a
subhalo at around 2 percent of the virial radius is vanishingly
small. Instead, we will consider only the uncertainty in the
properties of local substructure due to uncertainties in the
semi-analytic model. Fig.\ \ref{fig:11} shows the cumulative mass
function of objects in the solar neighbourhood (top panel), and the
cumulative contribution to the total mass in that volume. The dotted
line is for model A, and the solid line is for model B. We have
defined the solar neighbourhood as the region extending from 6 to
10\,kpc within our haloes, that is 2--3 percent of the virial
radius. The Virgo simulations have no substructure at all within an
equivalent volume with respect to the virial radius, and even in the
semi-analytic haloes substructure in this region is
rare. Nonetheless, we can get a sense of the mass function by
averaging over large numbers of trees.

We see that both the mass function and the mass fraction are very
sensitive to details of the model. The normalisation of mass fraction
in substructure changes by a factor of 1.5--2, but most of this
offset comes from the more massive haloes, which are comparatively
rare. There is also an appreciable change in the slope of the mass
function, however. This is worrying, as direct detection experiments
would be sensitive to irregularities on much smaller mass scales than
considered here. If we extrapolate assuming model `A', we estimate
that roughly 1 percent of mass in the solar neighbourhood would be in
substructure of $10^{6} M_{\odot}$ or more, and that 10 percent would
be in substructure of $10^{3} M_{\odot}$ or more. Extrapolating
naively in model `B', on the other hand, most of the mass of the halo
could be locked up in fairly massive subhaloes 
($M_{\rm s} \simgt 10^{4} M_{\odot}$). 
These estimates are very unreliable, but they
illustrate the fact that on solar mass scales (or spatial scales of
roughly a parsec), the distribution of dark matter could be extremely
irregular indeed. If local distribution is genuinely this lumpy, then
the limits placed by current experiments may be weakened considerably.
(Green 2003; Morgan et al.\ 2004). We will examine the fine structure
of local dark matter in detail in future work.

\section{Conclusion}\label{sec:6}

In this paper, we have compared the properties of halo substructure
predicted by a semi-analytic model with the substructure identified 
in a set of self-consistent numerical simulations of halo formation. 
The semi-analytic model combines merger trees, an algorithm for treating 
higher-order substructure, and an analytic description of satellite dynamics. 
While the original treatment of satellite dynamics had several free 
parameters (TB01), in our full model of halo formation these were fixed 
by comparison with the high-resolution, restricted simulations of 
Velazquez and White (1999) and H03, as described in paper I. Thus we 
have no remaining parametric freedom when comparing the predictions 
of the semi-analytic model to the numerical results. Our model does 
make a number of assumptions and approximations, however, concerning 
the shape and spherical symmetry of the halo density profile, for instance,
as well as halo concentrations and subhalo orbits. As discussed in 
paper II, we estimate that modifying these assumptions would change
our results at the 20--30 percent level.

Despite its uncertainties and simplifications, without any adjustment 
of the parameters the semi-analytic model does an excellent job of 
matching the numerical results in the outer regions of haloes, where 
the latter are the most robust. In particular, it matches the overall 
distribution of subhalo properties and the amplitude of the cumulative 
distributions of subhalo mass or peak velocity to within 10--20 percent. 
This is both the level of accuracy expected of our dynamical model, and 
is also comparable to the intrinsic scatter from one halo to the next, 
so we conclude that the two methods agree more or less exactly in this regime.

In contrast to this, in the central regions of haloes the
semi-analytic model predicts substantially more substructure than is seen
in the simulations. The excess subhaloes are predicted to be ancient,
dense systems which have orbited in the central part of the halo for
most of the age of the universe (10--12 Gyr, or since a redshift of
$z = 2$). In the semi-analytic model, these systems survive because
the overall heating and disruption rates are lower than those measured
in cosmological simulations. Whether these ancient systems should survive 
in reality is unclear. The disagreement with semi-analytic and numerical 
predictions could indicate that our dynamical model for heating and mass loss
is systematically less accurate. This seems unlikely, however, since the model 
is calibrated on the simpler, higher-resolution simulations of Velazquez and 
White (1999) and H03, and since many of the central subhaloes retain 20--30
percent of their original mass, and thus are at an evolutionary stage
where the analytic mass-loss model matches the restricted simulations 
quite closely.

One physical process recently proposed to explain the higher
disruption rate seen in simulations is the increased efficiency 
of mass loss in systems with anisotropic (internal) velocity
distributions (Kazantzidis et al.\ 2004). Our model is calibrated
using simulations of isotropic systems, so in principle anisotropy 
could affect our results. In the example they consider, however,
Kazantzidis et al.\ find that both fairly strong anisotropy and
substantial (80--90 percent) mass loss are required before the
evolution of the satellite changes substantially. Further work 
should clarify the importance of this effect.

The other possibility is that the semi-analytic predictions are
essentially correct, and that the simulations analysed in this work
are affected by residual overmerging, or by problems with the group
finder used to analyse their structure. If overmerging is responsible
for the patterns discussed in section \ref{sec:4}, higher-resolution
numerical work will be required to establish the true level of
substructure in the centres of CDM haloes definitively. In
particular, based on the arguments of section \ref{subsec:4.5}, the
properties of substructure in current simulations may be unreliable
within the central 10 percent of the virial radius (or 30\,kpc for a
system like the Milky Way), and it may take an increase of $\sim$100
or more in mass resolution to get convergent results down to the
equivalent of the solar radius. A similar increase in resolution would
be required to increase the relaxation time in a subhalo at the
nominal resolution limit of current simulations 
($\sim 10^{-5} M_{\rm vir,m}$), and that formed at $z = 2$, until 
it was longer than the Hubble time. Thus while future numerical work 
can eventually resolve this issue definitively, achieving the required 
mass and force resolution will remain challenging for some time.

Whatever the final answer to the problem, we have illustrated through
several examples that the survival of substructure in the innermost
parts of haloes is extremely important to the analysis of many
recent observational and experimental results. As it stands, our
semi-analytic model provides a robust and computationally efficient
basis for studying a wide range of problems related to halo structure
and substructure, including the origin and evolution of galaxy
morphology, tidal disruption of dwarf galaxies and globular clusters,
direct detection of dark matter particles and the local phase-space
density of dark matter, indirect detection of dark matter decay
products such as gamma-rays and positrons, and the analysis of
strong-lensing systems. We will explore these topics in future papers.

\section*{Acknowledgements}

The authors wish to thank E. Hayashi, S. Ghigna, B. Moore, J. Navarro
and T. Quinn for providing data from their simulations for comparison
with our model. We also wish to thank E. Hayashi, T.\ Kolatt, A.\
Kravtsov, J.\ Navarro, J. Silk, and S.\ White for helpful discussions.
JET gratefully acknowledges the support of a postgraduate scholarship
from the Natural Sciences \& Engineering Research Council of Canada
(NSERC) during the initial stages of this work, and support from the
Leverhulme Trust and from the UK Particle Physics and Astronomy
Research Council (PPARC) in the latter stages. AB gratefully
acknowledges support from NSERC through the Discovery and the
Collaborative Research Opportunities (CRO) grant programs.


\onecolumn

\begin{center}
{ 
Table I: Numerical Simulations

\begin{tabular}{lcccccccccccc}
\hline\hline
name&output&virial mass&virial radius&particle mass&softening&number of&$M_{\rm vir,m}/m_{\rm p}$&$r_{\rm s}/r_{\rm vir,m}$&references\\
 &redshift&$M_{\rm vir,m}$&$r_{\rm vir,m}$&$m_{\rm p}$&$r_{\rm s}$&subhaloes&(millions)&(\%)& \\
 & &(M$_{\odot}$)&(kpc)&(M$_{\odot}$)&(kpc)&$r < r_{\rm vir,m}$& & & \\
\hline
Coma& 0.0 & $2.37 \times 10^{15}$ & 3580 & $8.6 \times 10^{8}$ & 10.0 & 2302 & 2.76 & 0.28 & M98 \\
Virgo I& 0.0 & $4.1 \times 10^{14}$ & 1995 & $8.6 \times 10^{8}$ & 10.0 & 295 & 0.48 & 0.5 & M98 \\
Virgo IIa& 0.0 & $4.3 \times 10^{14}$ & 2026 & $1.1 \times 10^{8}$ & 1.0 & 1110 & 4.00 & 0.049 & G98,G00 \\
Virgo IIb& 0.1 & $3.98 \times 10^{14}$ & 1795 & $1.1 \times 10^{8}$& 1.0 & 1052 & 3.71 & 0.056 & G98,G00 \\
Andromeda& 0.2 & $2.12 \times 10^{12}$ & 288 & $2 \times 10^{6}$ & 1.5 & 250 & 1.06 & 0.52 & M99a,M99b \\
Milky Way& 0.2 & $1.59 \times 10^{12}$ & 261 & $2 \times 10^{6}$ & 1.5 & 280 & 0.80 & 0.57 & M99a,M99b \\
\hline
\end{tabular}
}
\end{center}


\begin{thebibliography}{}

\bibitem[Akerib et al.\ (2004)]{CDMS04} Akerib, D.\ S., et al.\ 
[The CDMS Collaboration] 2004, Phys.\ Rev.\ Lett., submitted (astro-ph/0405033)

\bibitem[Benoit et al.\ (2004)]{Edel04} Benoit, A.\ et al.\ 
[The EDELWEISS Collaboration] 2004, Phys.\ Rev.\ Lett. B, 545 43-49

\bibitem[Bernabei(2000)]{Bernabei:2000qi} Bernabei, R.\ et al.\ 
[The DAMA Collaboration] 2000, Phys.\ Rev.\ Lett. B, 480, 23

\bibitem[Bernabei(2003)]{Bernabei:2003qi} Bernabei, R.\ et al.\ 
[The DAMA Collaboration] 2003, Riv.\ Nuovo Cim., 26N1, 1

\bibitem[Chen, Kravtsov, \& Keeton(2003)]{2003ApJ...592...24C} Chen, J., 
Kravtsov, A.~V., \& Keeton, C.~R.\ 2003, \apj, 592, 24 

\bibitem[Dalal \& Kochanek(2002)]{2002ApJ...572...25D} Dalal, N.~\& 
Kochanek, C.~S.\ 2002, \apj, 572, 25 

\bibitem[De Lucia et al.(2004)]{2004MNRAS.348..333D} De Lucia, G., 
Kauffmann, G., Springel, V., White, S.~D.~M., Lanzoni, B., Stoehr, F., 
Tormen, G., \& Yoshida, N.\ 2004, MNRAS, 348, 333 

\bibitem[Desai et al.(2004)]{2004MNRAS.351..265D} Desai, V., Dalcanton, 
J.~J., Mayer, L., Reed, D., Quinn, T., \& Governato, F.\ 2004, \mnras, 351, 
265 

\bibitem[Diemand, Moore, Stadel, \& Kazantzidis(2004a)]{2004MNRAS.348..977D} 
Diemand, J., Moore, B., Stadel, J., \& Kazantzidis, S.\ 2004, \mnras, 348, 977
 
\bibitem[Diemand, Moore, \& Stadel(2004b)]{2004MNRAS.tmp..290D} Diemand, J., 
Moore, B., \& Stadel, J.\ 2004, \mnras, 290 

\bibitem[Diemand, Moore, \& Stadel(2004c)]{2004MNRAS.352..535D} Diemand, J., 
Moore, B., \& Stadel, J.\ 2004, \mnras, 352, 535 

\bibitem[D'Onghia \& Lake (2004)]{dong04}D'Onghia, E., Lake, G.\ 2004, 
ApJ, in press (astro-ph/0309735)

\bibitem[Eke, Navarro, \& Steinmetz(2001)]{2001ApJ...554..114E} Eke, V.~R., 
Navarro, J.~F., \& Steinmetz, M.\ 2001, ApJ, 554, 114 (ENS01)

\bibitem[Evans \& Witt(2003)]{2003MNRAS.345.1351E} Evans, N.~W.~\& Witt, 
H.~J.\ 2003, \mnras, 345, 1351 

\bibitem[Gao, De Lucia, White, \& Jenkins (2004a)]{2004MNRAS.352L...1G} Gao, 
L., De Lucia, G., White, S.~D.~M., \& Jenkins, A.\ 2004, \mnras, 352, L1

\bibitem[Gao et al.\ (2004b)]{gao2} Gao, L., White, S.~D.~M., 
Jenkins, A., Stoehr, F., \& Springel, V.\ 2004, \mnras, submitted
(astro-ph/0404589)

\bibitem[Gavazzi et al.(2004)]{2004A&A...422..407G} Gavazzi, R., Mellier, 
Y., Fort, B., Cuillandre, J.-C., \& Dantel-Fort, M.\ 2004, \aap, 422, 407 

\bibitem[Ghigna et al.(1998)]{1998MNRAS.300..146G} Ghigna, S., Moore, B., 
Governato, F., Lake, G., Quinn, T., \& Stadel, J.\ 1998, MNRAS, 300, 146 (G98) 

\bibitem[Ghigna et al.(2000)]{2000ApJ...544..616G} Ghigna, S., Moore, B., 
Governato, F., Lake, G., Quinn, T., \& Stadel, J.\ 2000, ApJ, 544, 616 (G00)

\bibitem[Gill, Knebe, \& Gibson(2004a)]{2004MNRAS.351..399G} Gill, S.~P.~D., 
Knebe, A., \& Gibson, B.~K.\ 2004a, \mnras, 351, 399 

\bibitem[Gill, Knebe, Gibson, \& Dopita(2004b)]{2004MNRAS.351..410G} Gill, 
S.~P.~D., Knebe, A., Gibson, B.~K., \& Dopita, M.~A.\ 2004b, \mnras, 351, 410 

\bibitem[Governato, Ghigna, Moore(2001)]{2001aats.conf..469G} 
Governato F., Ghigna S., Moore B., 2001, 
in ASP Conf.~Ser.~245: Astrophysical Ages and Times Scales, 469 

\bibitem[Green (2003)]{green03}Green, A.\ M.\ 2003, 
Phys.\ Rev.\ D, 68, 023004 

\bibitem[Hayashi et al.(2003)]{2003ApJ...584..541H} Hayashi, E., Navarro, 
J.~F., Taylor, J.~E., Stadel, J., \& Quinn, T.\ 2003, ApJ, 584, 541 (H03)

\bibitem[Helmi, White, \& Springel(2003)]{2003MNRAS.339..834H} Helmi, A., 
White, S.~D.~M., \& Springel, V.\ 2003, \mnras, 339, 834 

\bibitem[Inoue \& Chiba(2003)]{2003ApJ...591L..83I} Inoue, K.~T.~\& Chiba, 
M.\ 2003, \apjl, 591, L83 

\bibitem[Kauffmann, Colberg, Diaferio, \& White(1999)]{1999MNRAS.303..188K} 
Kauffmann, G., Colberg, J.~M., Diaferio, A., \& White, S.~D.~M.\ 1999, 
MNRAS, 303, 188 

\bibitem[Kazantzidis et al.(2004)]{2004ApJ...608..663K} Kazantzidis, S., 
Mayer, L., Mastropietro, C., Diemand, J., Stadel, J., \& Moore, B.\ 2004, 
\apj, 608, 663 

\bibitem[Khochfar and Burkert (2003)]{khoch} Khochfar, S., \& 
Burkert, A.\ 2003, MNRAS, submitted (astro-ph/0309611)

\bibitem[Kim et al.(2004)]{2004MNRAS.347..355K} Kim, T.-S., Viel, M., 
Haehnelt, M.~G., Carswell, R.~F., \& Cristiani, S.\ 2004, MNRAS, 347, 355 

\bibitem[Klypin, Gottl{\" o}ber, Kravtsov, \& Khokhlov(1999a)]
{1999ApJ...516..530K} Klypin, A., Gottl{\" o}ber, S., 
Kravtsov, A.~V., \& Khokhlov, A.~M.\ 1999, ApJ, 516, 530 

\bibitem[Klypin, Kravtsov, Valenzuela, \& Prada(1999b)]{1999ApJ...522...82K} 
Klypin, A., Kravtsov, A.~V., Valenzuela, O., \& Prada, F.\ 1999, ApJ, 522, 
82 

\bibitem[Liubarsky et al.\ (2000)]{liub}
Liubarsky, I., et al.\ 2000, Nucl. Phys. B, (Proc. Suppl.), 87, 64

\bibitem[Mathis et al.(2002)]{2002MNRAS.333..739M} Mathis, H., Lemson, G., 
Springel, V., Kauffmann, G., White, S.~D.~M., Eldar, A., \& Dekel, A.\ 
2002, \mnras, 333, 739 

\bibitem[Metcalf (2004)]{met04} Metcalf, R. B.\ 2004, apJ, submitted
(astro-ph/0407298)

\bibitem[Metcalf, Moustakas, Bunker, \& Parry(2004)]{2004ApJ...607...43M} 
Metcalf, R.~B., Moustakas, L.~A., Bunker, A.~J., \& Parry, I.~R.\ 2004, 
\apj, 607, 43 

\bibitem[Moore, Katz, \& Lake(1996)]{1996ApJ...457..455M} Moore, B., Katz, 
N., \& Lake, G.\ 1996, \apj, 457, 455 

\bibitem[Moore et al.(1996)]{1996Natur.379..613M} Moore, B., Katz, N., 
Lake, G., Dressler, A., \& Oemler, A.\ 1996, Nature, 379, 613 

\bibitem[Moore et al.(1998)]{1998ApJ...499L...5M} Moore, B., Governato, F.,
Quinn, T., Stadel, J., \& Lake, G.\ 1998, ApJ, 499, L5 (M98)

\bibitem[Moore et al.(1999a)]{metal99} Moore B., Ghigna S., Governato F., 
Lake G., Quinn T., Stadel J., \& Tozzi P., 1999, ApJ, 524, L19 (M99a)

\bibitem[Moore et al.(1999b)]{1999MNRAS.310.1147M} Moore, B., Quinn, T., 
Governato, F., Stadel, J., \& Lake, G.\ 1999, MNRAS, 310, 1147 (M99b)

\bibitem[Morgan, Green, \&Spooner (2004)]{mgs04} Morgan, B., Green, A. M.,
\& Spooner, N.\ J.\ C.\ 2004, preprint (astro-ph/0408047)

\bibitem[Moustakas \& Metcalf(2003)]{2003MNRAS.339..607M} Moustakas, 
L.~A.~\& Metcalf, R.~B.\ 2003, \mnras, 339, 607 

\bibitem[Nagai \& Kravtsov(2004)]{nagai04}
Nagai, D., Kravtsov, A. V.\ 2004, ApJ, submitted (astro-ph/0408273)

\bibitem[Natarajan, Kneib, \& Smail(2002)]{2002ApJ...580L..11N} Natarajan, 
P., Kneib, J., \& Smail, I.\ 2002, ApJL, 580, L11 

\bibitem[Okamoto \& Habe(1999)]{1999ApJ...516..591O} Okamoto, T.~\& Habe, 
A.\ 1999, \apj, 516, 591 

\bibitem[Pretzl(2002)]{2002SSRv..100..209P} Pretzl, K.\ 2002, Space Science 
Reviews, 100, 209 

\bibitem[Reed et al.(2003)]{2003MNRAS.346..565R} Reed, D., Gardner, J., 
Quinn, T., Stadel, J., Fardal, M., Lake, G., \& Governato, F.\ 2003, 
MNRAS, 346, 565 

\bibitem[Reed et al.(2004)]{reed04}Reed, D., Governato, F., Quinn, T., 
Gardner, J., Stadel, J., Lake, G.\ 2004, \mnras, submitted (astro-ph/0406034)

\bibitem[Rhodes et al.(2004)]{2004ApJ...605...29R} Rhodes, J., Refregier, 
A., Collins, N.~R., Gardner, J.~P., Groth, E.~J., \& Hill, R.~S.\ 2004, 
\apj, 605, 29 

\bibitem[Schechter \& Wambsganss(2002)]{2002ApJ...580..685S} Schechter, 
P.~L.~\& Wambsganss, J.\ 2002, \apj, 580, 685 

\bibitem[Spergel et al.(2003)]{2003ApJS..148..175S} 
Spergel, D.~N.~et al.\ 2003, ApJS, 148, 175

\bibitem[Springel, White, Tormen, \& Kauffmann(2001)]{2001MNRAS.328..726S} 
Springel, V., White, S.~D.~M., Tormen, G., \& Kauffmann, G.\ 2001, MNRAS, 
328, 726 

\bibitem[Stadel(2001)]{std01}Stadel, J.\ 2001, Ph.D. thesis, University
of Washington

\bibitem[Stoehr, White, Tormen, \& Springel(2002)]{2002MNRAS.335L..84S} 
Stoehr, F., White, S.~D.~M., Tormen, G., \& Springel, V.\ 2002, MNRAS, 
335, L84 

\bibitem[Taylor (2002)]{t2001}Taylor, J.~E.\ 2002, Ph.D. thesis, University 
of Victoria \\
\hspace{0.5truein} (http://wwwlib.umi.com/dissertations/fullcit/NQ62530)

\bibitem[Taylor \& Babul(2001)]{2001ApJ...559..716T} Taylor, J.~E.~\& 
Babul, A.\ 2001, ApJ, 559, 716 

\bibitem[Taylor \& Babul(2004a)]{2004MNRAS.348..811T} Taylor, J.~E.~\& 
Babul, A.\ 2004, MNRAS, 348, 811 

\bibitem[Taylor \& Babul(2004b)]{paperIII} Taylor, J.~E.~\& 
Babul, A.\ 2004b, MNRAS, submitted (paper III)

\bibitem[Taylor \& Navarro(2001)]{2001ApJ...563..483T} Taylor, J.~E., \& 
Navarro, J.~F.\ 2001, ApJ, 563, 483 

\bibitem[Tegmark et al.(2004)]{2004ApJ...606..702T} Tegmark, M., et al.\ 
2004, \apj, 606, 702 

\bibitem[van Kampen(1995)]{1995MNRAS.273..295V} van Kampen, E.\ 1995, 
\mnras, 273, 295 

\bibitem[Velazquez \& White(1999)]{1999MNRAS.304..254V} Velazquez, H.~\& 
White, S.~D.~M.\ 1999, \mnras, 304, 254 

\bibitem[Weller, Ostriker, \& Bode]{wel04} Weller, J., Ostriker J.\ P., 
\& Bode, P.\ 2004, \mnras, submitted (astro-ph/0405445) 

\bibitem[Yonehara, Umemura, \& Susa(2003)]{2003PASJ...55.1059Y} Yonehara, 
A., Umemura, M., \& Susa, H.\ 2003, \pasj, 55, 1059 

\bibitem[Zentner \& Bullock(2003)]{2003ApJ...598...49Z} Zentner, A.~R.~\& 
Bullock, J.~S.\ 2003, \apj, 598, 49 

\end{thebibliography}
\end{document}